\documentclass[11pt,a4paper]{article}

\usepackage{jheppub}

\usepackage{slashed}

\DeclareMathOperator{\Tr}{Tr}

\def\II{\hbox{{1}\kern-.25em\hbox{l}}}

\newcommand \vev [1] {\langle{#1}\rangle}

\newcommand{\uu}{{\uparrow\!\uparrow}}
\newcommand{\dd}{{\downarrow\!\downarrow}}
\newcommand{\ud}{{\uparrow\!\downarrow}}
\newcommand{\du}{{\downarrow\!\uparrow}}

\newcommand{\derleft}{\stackrel{\leftarrow}{D}}

\newcommand \widebar [1] {\overline{#1}}

\def \e {\mbox{e}}

\title{Higher-twist B-meson Distribution Amplitudes in HQET}

\author[a]{V. M. Braun}

\author[a]{Yao Ji}

\author[b,a]{and A. N. Manashov}

\subheader{\large\textsc{DESY~17-037}}

\affiliation[a]{
   Institut f\"ur Theoretische Physik, Universit\"at
   Regensburg \\ D-93040 Regensburg, Germany}

\affiliation[b]{
   Institut f\"ur Theoretische Physik, Universit\"at Hamburg\\
   D-22761 Hamburg, Germany}

\emailAdd{vladimir.braun@physik.ur.de}

\emailAdd{yao.ji@ur.de}

\emailAdd{alexander.manashov@desy.de}

\abstract{
We present a systematic study of higher-twist distribution amplitudes (DAs) of the B-meson which
give rise to power-suppressed $1/m_B$ contributions to B-decays in final states with energetic
light particles in the framework of QCD factorization. As the main result, we find that the renormalization group equations
for the three-particle distributions are completely integrable in the large $N_c$ limit and can be solved exactly.
General properties of the solutions are studied. We propose simple models
for higher-twist DAs which satify all existing constraints and can be used in phenomenological studies.
}


\begin{document}

\maketitle

\newpage

%
\section{Introduction}
%

B-meson light-cone distribution amplitudes (DAs) are the main nonperturbative input to the
QCD description of weak decays involving light hadrons in the framework of
QCD factorization~\cite{Beneke:1999br,Beneke:2000ry,Beneke:2000wa}.
The simplest light antiquark-heavy quark DA gives the dominant contribution in the expansion in powers of the heavy quark mass
and received much attention
already~\cite{Grozin:1996pq,Lange:2003ff,Braun:2003wx,Lee:2005gza,Bell:2013tfa,Braun:2014owa,Feldmann:2014ika}.
It has become increasingly clear, however, that the leading power accuracy is not sufficient and
the power of the QCD factorization approach depends crucially on the possibility to control, or at least estimate,
the corrections suppressed by powers of the $b$-quark mass. This task is very challenging due to
infrared divergences which appear in power-suppressed contributions in the purely perturbative framework.
In recent years there had been some progress in this direction based on the combination of light-cone sum rule approach
with the expansion in terms of B-meson DAs~\cite{Khodjamirian:2006st,DeFazio:2007hw,Braun:2012kp,Wang:2015vgv,Wang:2016beq,Wang:2017jow}.
In this technique the so-called soft or end-point nonfactorizable contributions to B-decays can be calculated in
terms of the DAs of increasing twist.  One of the problems on this way is that higher-twist B-meson DAs involve
contributions of multiparton states and are practically unknown.

In the recent paper~\cite{Braun:2015pha} it was pointed out that the structure of subleading twist-three B-meson DAs
is simpler than expected.  In particular the twist-three DA  $\phi_-(\omega)$ evolves autonomously with the scale and
does not mix with ``genuine'' three-particle contributions. In this work we extend this analysis to twist-four
DAs. Our main result is that the corresponding renormalization group equations (RGE) are completely integrable and can be
solved exactly. Combined with the relations that follow from QCD equations of motion, this structure provides one with
a set of robust constraints on the DAs and allows one to build phenomenologically acceptable models with a minimum number
of free parameters.

The presentation is organized as follows. In Sect.~\ref{sec:general} we present the general classification of the existing two-
and three-particle B-meson DAs and discuss the corresponding twist and conformal spin assignment. We also use this
Section to introduce our notation. Sect.~\ref{sec:scale} contains our main results. We discuss here the  scale dependence 
of the three-particle twist-four DAs, find explicit analytic solution  
and work out the representation of the DAs in terms of the eigenfunctions of the evolution kernels. 
The derivation uses the formalism of the Quantum Inverse Scattering Method (QISM)~\cite{Faddeev:1979gh,Kulish:1981bi,Faddeev:1996iy} 
and is sketched
in Appendix~\ref{app:RGequations};  details will be published elsewhere. 
The two-particle higher-twist DAs that appear in the light-cone expansion of the heavy-light correlation functions
are discussed in Sect.~\ref{sec:KKQT}, based on the earlier work by Kawamura \emph{et al.}~(KKQT)~\cite{Kawamura:2001jm}.
The relations between different DAs due to QCD equations of motion (EOM) are discussed in detail. 
Two new relations between the three-particle DAs are derived in App.~\ref{app:WW} using a different technique based on the 
two-component spinor formalism. Several simple models for the higher-twist DAs that satisfy EOM constraints 
are introduced and discussed shortly in Sect.~\ref{sec:Models}. The final Sect.~\ref{sec:Summary} contains a summary
of our results and a short discussion of the remaining problems.

%
\section{General classification}\label{sec:general}
%

Following~\cite{Grozin:1996pq} we define the B-meson DAs
as matrix elements of the renormalized nonlocal operators built of an effective heavy quark field $h_v(0)$ and
a light (anti)quark at a light-like separation:
\begin{align}
 \langle 0| \bar q(nz) \Gamma [nz,0] h_v(0) |\bar B(v)\rangle &=
-\frac{i}2  F_B(\mu)\! \Tr\Big\{\!\gamma_5 \Gamma P_+ \Big[ \Phi_+(z,\mu)
-\frac12\slashed{n} \Big(\Phi_+(z,\mu)\!-\!\Phi_-(z,\mu)\Big)\Big]\!\Big\},
\label{def:two}
\end{align}
where
\begin{equation}\label{WL}
 {}[zn,0] \equiv {\rm Pexp}\left[ig\int_0^1\!du\,n_\mu A^\mu(uzn)\right]
\end{equation}
is the Wilson line factor that ensures gauge invariance. Such factors are always implied but will often be omitted
for brevity.

Here and below  $v_\mu$ is the heavy quark velocity,
$n_\mu$ is the light-like vector, $n^2=0$, such that $n\cdot v=1$,
$P_+=\frac12(1+\not\!v)$, $\Gamma$ stands for an arbitrary Dirac structure,
$|\bar B(v)\rangle$ is the $\bar B$-meson state in the heavy quark
effective theory (HQET) and $F_B(\mu)$ is the HQET decay constant
which is related to the physical B-meson decay constant,
to one-loop accuracy, as
\begin{equation}
\label{defF}
   f_B\sqrt{m_B} = F_B(\mu)\left[1+\frac{C_F\alpha_s}{4\pi}\!
   \left(3\ln\frac{m_b}{\mu}-2\right)+\ldots\right].
\end{equation}
The parameter $z$ specifies the light (anti)quark position on the light cone.
To fix the normalization, we assume
\begin{align}
   n_\mu = (1,0,0,1)\,, && \bar n_\mu = (1,0,0-1)\,, && v_\mu = \frac12 (n_\mu + \bar n_\mu)\,, && (n\cdot \bar n) =2\,.
\label{nbarn}
\end{align}
The functions $\Phi_+(z,\mu)$ and $\Phi_-(z,\mu)$ are the leading- and subleading-twist
two-particle B-meson DAs~\cite{Beneke:2000wa}. They are analytic functions of $z$
in the lower half-plane, $\text{Im}(z)<0$, and are related by Fourier transform to the momentum space DAs
\begin{align}
 \Phi_\pm(z,\mu) &= \int\limits_0^\infty d\omega \, e^{-i\omega z}\phi_\pm(\omega,\mu)\,.
\end{align}
The inverse transformation reads
\begin{align}
\phi_\pm(\omega,\mu) & = \frac{1}{2\pi} \int\limits_{-\infty}^\infty dz \,  e^{i\omega z} \Phi_\pm(z-i\epsilon,\mu)\,.
\end{align}
Note  that we use upper case letters for the coordinate-space and low case for the momentum-space distributions.

The three-particle quark-gluon DAs are more numerous. There exist eight independent Lorentz structures~\cite{Geyer:2005fb}
 and therefore eight invariant functions that can be defined as
\begin{eqnarray}
\lefteqn{\langle 0| \bar q(nz_1) gG_{\mu\nu}(nz_2)\Gamma h_v(0) |\bar B(v)\rangle =}
\nonumber\\
&=&
\frac12 F_B(\mu) \Tr\biggl\{\gamma_5 \Gamma P_+
\biggl[ (v_\mu\gamma_\nu-v_\nu\gamma_\mu)  \big[{\Psi}_A-{\Psi}_V \big]-i\sigma_{\mu\nu}{\Psi}_V
- (n_\mu v_\nu-n_\nu v_\mu){X}_A
\nonumber\\&&{}\hspace*{0.1cm}
 + (n_\mu \gamma_\nu-n_\nu \gamma_\mu)\big[W+{Y}_A\big]
- i\epsilon_{\mu\nu\alpha\beta} n^\alpha v^\beta \gamma_5 \widetilde{X}_A
+ i\epsilon_{\mu\nu\alpha\beta} n^\alpha \gamma^\beta\gamma_5 \widetilde{Y}_A
\nonumber\\&&{}\hspace*{0.1cm}
- (n_\mu v_\nu-n_\nu v_\mu)\slashed{n}\,{W} + (n_\mu \gamma_\nu-n_\nu \gamma_\mu)\slashed{n}\,{Z}
\biggr]\biggr\}(z_1,z_2;\mu)\,.
\label{def:three}
\end{eqnarray}

Our notation follows, where possible, the original definition in Ref.~\cite{Kawamura:2001jm}.
We use the standard Bjorken-Drell convention \cite{BD65} for the metric and the Dirac matrices; in particular
$\gamma_5 = i \gamma^0\gamma^1\gamma^2\gamma^3$, and the Levi-Civita tensor $\epsilon_{\alpha\beta\mu\nu}$
defined as a totally antisymmetric tensor with $\epsilon_{0123} =1$. The covariant derivative is defined as
$D_\mu = \partial_\mu -igA_\mu$ and the dual gluon strength tensor as
$\widetilde{G}_{\mu\nu} = \frac12 \epsilon_{\mu\nu\alpha\beta} G^{\alpha\beta}$.
The momentum space distributions are defined as
\begin{align}
 {\Psi}_A (z_1,z_2) &=
\int_0^\infty \!\!d\omega_1\!  \int_0^\infty \!\!d\omega_2\,\, e^{-i\omega_1 z_1-i\omega_2 z_2}\, {\psi}_A (\omega_1,\omega_2)
\label{3pt-momspace}
\end{align}
and similarly for the other functions.

In practical calculations the gluon field strength tensor is often contracted with the light-like vector.
The definition in \eqref{def:three} leads to the following pair of equations:
\begin{eqnarray}
\lefteqn{\langle 0| \bar q(z_1n) gG_{\mu\nu}(z_2n) n^\nu \Gamma h_v(0) |\bar B(v)\rangle=}
\nonumber\\ &=&
\frac12 F_B(\mu) \Tr\biggl\{\gamma_5 \Gamma P_+\biggl[
(\slashed{n}v_\mu - \gamma_\mu) \big({\Psi}_A-{\Psi}_V \big)
-i \sigma_{\mu\nu} n^\nu  {\Psi}_V - n_\mu {X}_A + n_\mu  \slashed{n} {Y}_A \biggr] \biggr\}(z_1,z_2;\mu)\,,
\nonumber\\
\lefteqn{\langle 0| \bar q(z_1n) ig\widetilde{G}_{\mu\nu}(z_2n) n^\nu \gamma_5 \Gamma h_v(0) |\bar B(v)\rangle=}
\nonumber\\ &=&
\frac12 F_B(\mu) \Tr\biggl\{\gamma_5 \Gamma P_+\biggl[
(\slashed{n}v_\mu - \gamma_\mu) \big(\widetilde{\Psi}_A-\widetilde{\Psi}_V \big)
-i \sigma_{\mu\nu} n^\nu  \widetilde{\Psi}_V - n_\mu \widetilde{X}_A + n_\mu  \slashed{n} \widetilde{Y}_A\biggr]\biggr\}(z_1,z_2;\mu)\,,
\nonumber\\
\end{eqnarray}
cf.~\cite{Kawamura:2001jm}.
It is easy to show that
\begin{align}
   \widetilde{\Psi}_A = - \Psi_V\,, \qquad  \widetilde{\Psi}_V = - \Psi_A\,,
\end{align}
but  the $X, Y$- and $\widetilde{X}, \widetilde{Y}$- functions are not related to each other.
Note that the DAs $W$ and $Z$ do not appear in these expressions.

In what follows we will use shorthand notations for the field coordinates on the light cone:
$$
   \bar q(z_1) \equiv \bar q(z_1 n)\,, \qquad\qquad G_{\mu\nu}(z_2) \equiv G_{\mu\nu}(z_2 n)\,.
$$

%
\subsection{Collinear twist and conformal spin assignment}
%

The basis of the DAs in \eqref{def:three} is convenient because of the simple Lorentz structures. However, it is not
suitable for discussion of QCD factorization as the DAs in this basis do not have definite (collinear) twist.
Hence terms with different power suppression in the heavy quark expansion get mixed.

Twist $t$ and conformal spin $j$ of the light quark and gluon fields are given by the usual expressions
\begin{align}
   t = d - s\,,  \qquad j = \frac12(d+s)\,,
\end{align}
where $d$ is the canonical dimension and $s$ is the spin \emph{projection} on the light cone.
Twist of a nonlocal heavy-light operator can then be defined as the sum of twists of
the light constituents plus one unit of twist for the effective heavy quark field  $h_v$.
Adding one unit of twist for $h_v$ is entirely a convention which we adopt in order to match the usual
twist hierarchy for light quark-gluon operators; in this way in both cases the leading-twist
contributions are defined as twist-two.

The DAs of definite twist and conformal spins of the constituent fields are
most easily defined by the corresponding projections of the general expression in \eqref{def:three}.
One finds one DA of twist three
\begin{align}
    2 F_B(\mu) \Phi_3(z_1,z_2;\mu) &= \langle 0| \bar q(z_1) gG_{\mu\nu}(z_2) n^\nu \slashed{n}\gamma_\perp^\mu \gamma_5 h_v(0) |\bar B(v)\rangle\,,
\end{align}
where
\begin{align}
  \Phi_3 &= {\Psi}_A-{\Psi}_V\,,
\end{align}
three independent twist-four DAs
\begin{align}
    2 F_B(\mu) \Phi_4(z_1,z_2;\mu) &=
\langle 0| \bar q(z_1) gG_{\mu\nu}(z_2) n^\nu \slashed{\bar n} \gamma^\mu_\perp\gamma_5 h_v(0) |\bar B(v)\rangle\,,
\notag\\[1mm]
    2 F_B(\mu) \Psi_4(z_1,z_2;\mu) &=
\langle 0| \bar q(z_1) gG_{\mu\nu}(z_2)\bar n^\mu n^\nu \slashed{n}\gamma_5 h_v(0) |\bar B(v)\rangle\,,
\notag\\[1mm]
    2 F_B(\mu) \widetilde{\Psi}_4(z_1,z_2;\mu) &=
\langle 0| \bar q(z_1) ig\widetilde{G}_{\mu\nu}(z_2)\bar n^\mu n^\nu \slashed{n} h_v(0) |\bar B(v)\rangle\,,
\end{align}
where
\begin{align}
  \Phi_4 &= {\Psi}_A+{\Psi}_V\,, 
\notag\\
  \Psi_4 &= {\Psi}_A+{X}_A\,,
\notag\\
  \widetilde{\Psi}_4 &=  {\Psi}_V- \widetilde{X}_A\,,
\end{align}
three twist-five DAs
\begin{align}
    2 F_B(\mu) \widetilde{\Phi}_5(z_1,z_2;\mu) &=
\langle 0| \bar q(z_1) gG_{\mu\nu}(z_2)\bar n^\nu \slashed{ n} \gamma^\mu_\perp\gamma_5 h_v(0) |\bar B(v)\rangle\,,
\notag\\[1mm]
    2 F_B(\mu) \Psi_5(z_1,z_2;\mu) &=
\langle 0| \bar q(z_1) gG_{\mu\nu}(z_2)\bar n^\mu n^\nu \slashed{\bar n}\gamma_5 h_v(0) |\bar B(v)\rangle\,,
\notag\\[1mm]
    2 F_B(\mu) \widetilde{\Psi}_5(z_1,z_2;\mu) &=
\langle 0| \bar q(z_1) ig\widetilde{G}_{\mu\nu}(z_2)\bar n^\mu n^\nu \slashed{\bar n} h_v(0) |\bar B(v)\rangle\,,
\end{align}
where
\begin{align}
\widetilde{\Phi}_5 & = \Psi_A +  \Psi_V + 2 Y_A - 2 \widetilde{Y}_A + 2 W\,,
\notag\\
  \Psi_5 &= -{\Psi}_A  +  {X}_A  -  2 Y_A\,,
\nonumber\\
  \widetilde{\Psi}_5 &= - {\Psi}_V  -  \widetilde{X}_A  + 2 \widetilde{Y}_A\,,
\end{align}
and one twist-six DA
\begin{align}
 2 F_B(\mu) \widetilde{\Phi}_6(z_1,z_2;\mu) &=
\langle 0| \bar q(n z_1) gG_{\mu\nu}(n z_2) \bar n^\nu \slashed{\bar n} \gamma^\mu_\perp\gamma_5 h_v(0) |\bar B(v)\rangle
\end{align}
with
\begin{align}
 \Phi_6 &= {\Psi}_A- {\Psi}_V + 2{Y}_A + 2{W}+ 2 \widetilde{Y}_A - 4  Z\,.
\end{align}
The conformal spin assignment for all DAs is summarized in Table.~\ref{table:1}.

The twist-five and twist-six DAs are not expected to contribute to the leading power corrections
$\mathcal{O}(1/m_B)$ in B-decays and will not be considered further in this work.

Note that we also do not consider twist-four four-particle B-meson DAs (with two gluon fields and/or
with an extra quark-antiquark pair). By analogy to the DAs of light mesons twist-four four-particle DAs are expected to
be small and, most importantly, have autonomous scale dependence i.e. they do not mix with the three-particle DAs.
Thus they can be consistently put to zero at all scales and do not reappear via evolution.

\begin{table}[t]
\renewcommand{\arraystretch}{1.2}
\begin{center}
\begin{tabular}{|c|c|c|c|c|c|c|c|c|}
\hline\hline
       & $\Phi_3$ &  $\Phi_4$ & $\Psi_4+ \widetilde{\Psi}_4$ & $\Psi_4-\widetilde{\Psi}_4 $ & $\Phi_5$ &  $\Psi_5+ \widetilde{\Psi}_5$ & $\Psi_5 - \widetilde{\Psi}_5$ &  ${\Phi}_6$
\\
\hline
twist&     $3$ &  $4$ & $4$ & $4$ & $5$ &  $5$ &  $5$ &  $6$
\\
\hline
$j_q$&     $1$ &  $1/2$ & $1$ & $1$ & $1$ &  $1/2$ &  $1/2$ &  $1/2$
\\
\hline
$j_g$&     $3/2$ &  $3/2$ & $1$ & $1$ & $1/2$ &  $1$ &  $1$  &   $1/2$
\\
\hline
chirality   & $\ud(\du)$ & $\uu(\dd)$ & $\uu(\dd)$ &  $\ud(\du)$ & $\uu(\dd)$ & $\uu(\dd)$ &  $\ud(\du)$ &   $\ud(\du)$
\\
\hline
\end{tabular}
\end{center}
\caption{\small The twist, conformal spins $j_q$, $j_g$ of the constituent fields
             and chirality [same  or opposite] of the three-particle B-meson DAs.
}
\label{table:1}
\renewcommand{\arraystretch}{1.0}
\end{table}

%
\subsection{Asymptotic behavior at small momenta}
%
Asymptotic behavior of all DAs at small quark and gluon momenta is determined by conformal
spins of the fields~\cite{Braun:1989iv}
\begin{align}
   f(\omega_1,\omega_2) \sim \omega_1^{2j_1-1}\omega_2^{2j_2-1}\,.\qquad \qquad f \in \{\phi_3, \phi_4,\psi_4,\tilde\psi_4\,\ldots\}
\end{align}
In particular
\begin{align}
 \phi_3(\omega_1,\omega_2) \sim \omega_1 \omega_2^2\,, \qquad \phi_4(\omega_1,\omega_2)\sim \omega_2^2\,,
\qquad \psi_4(\omega_1,\omega_2)\sim \widetilde{\psi}_4(\omega_1,\omega_2) \sim \omega_1\omega_2\,.
\label{small-momenta}
\end{align}
These expressions can be verified considering correlation functions of the corresponding light-ray
operators and suitable local currents, e.g.~\cite{Khodjamirian:2006st}, and are stable against evolution
provided the renormalization group equations respect conformal symmetry which is true to the leading
logarithmic accuracy.

%
\subsection{Spinor representation}\label{subsec:spinor}
%

Discussion of scale dependence of the DAs is considerably simplified using spinor formalism. In this work we follow
the conventions adopted in \cite{Braun:2008ia,Braun:2009vc}.

Any light-like vector can be represented by a product of two spinors. We write
\begin{align}
n_{\alpha\dot\alpha}=n_\mu\sigma^\mu_{\alpha\dot\alpha}=\lambda_\alpha\bar\lambda_{\dot\alpha}\,,
&& \bar n_{\alpha\dot\alpha}=\bar n_\mu\sigma^\mu_{\alpha\dot\alpha}=\mu_\alpha\bar\mu_{\dot\alpha}
\end{align}
where $\bar\lambda = \lambda^\dagger$, $\bar\mu = \mu^\dagger$.  We choose
\begin{align}
(\lambda\,\mu)=\lambda^\alpha \mu_\alpha=2, &&
(\bar\mu\, \bar\lambda)=\bar\mu_{\dot\alpha} \bar\lambda^{\dot\alpha}=2
\label{norm-spinors}
\end{align}
which is consistent with our normalization $(n\bar n) =2$.%
\footnote{The particular choice in Eq.~\eqref{nbarn} corresponds to $\lambda^\alpha=\sqrt2\,(0,1)$, $\mu^\alpha=\sqrt2\,(1,0)$.}

The ``+'' and ``--'' fields are defined as the projections onto the auxiliary $\lambda$ and $\mu$ spinors,
\begin{align}
\chi_+=\lambda^\alpha \psi_\alpha, && \bar\psi_+=\bar\lambda^{\dot \alpha} \psi_{\dot\alpha},
&& f_{++}=\lambda^\alpha\lambda^\beta f_{\alpha\beta}, && f_{+-}=\lambda^\alpha\mu^\beta f_{\alpha\beta}\,,
&&
\bar f_{++}=\bar\lambda^{\dot \alpha}\bar\lambda^{\dot\beta} \bar f_{\dot\alpha\dot\beta}
\end{align}
etc.
 The Dirac
(antiquark) spinor 
$$
 q=\begin{pmatrix}
\psi_\alpha\\\bar \chi^{\dot\beta}\end{pmatrix}\,, \qquad\qquad \bar q=(\chi^\beta,\bar\psi_{\dot\alpha}) 
$$
is written in this notation as
\begin{align}
(\lambda\mu)\,\chi^\alpha=\mu^\alpha \chi_+  - \lambda^\alpha \chi_-\,,
&&
(\bar \mu\bar\lambda)\,\bar \psi_{\dot\alpha}=\bar \mu_{\dot\alpha} \bar \psi_+  - \bar\lambda_{\dot\alpha}\bar \psi_-\,.
\end{align}
The equation of motion (EOM) for the effective heavy quark field $\slashed{v}h_v = h_v $ reads
\begin{align}
    h_{+} = - \bar h_-\,, \qquad h_- =  \bar h_+\,.
\label{EOMheavy}
\end{align}
The gluon strength tensor $F_{\mu\nu}$ can be decomposed as
\begin{align}
F_{\alpha\beta,\dot\alpha\dot\beta} =\sigma^\mu_{\alpha\dot\alpha} \sigma^\nu_{\beta\dot\beta} F_{\mu\nu}=
2\left(\epsilon_{\dot\alpha\dot\beta} f_{\alpha\beta}-
\epsilon_{\alpha\beta} \bar f_{\dot\alpha\dot\beta}
\right),
\notag\\
i {\widetilde F}_{\alpha\beta,\dot\alpha\dot\beta}=
\sigma^\mu_{\alpha\dot\alpha} \sigma^\nu_{\beta\dot\beta}i\widetilde F_{\mu\nu}=
2\left(\epsilon_{\dot\alpha\dot\beta}f_{\alpha\beta}+
\epsilon_{\alpha\beta}\bar f_{\dot\alpha\dot\beta}\right).
\end{align}
Here $f_{\alpha\beta}$ and $\bar f_{\dot\alpha\dot\beta}$ are chiral and antichiral
symmetric tensors, $f^*=\bar f$, which belong to $(1,0)$ and $(0,1)$ representations
of the Lorenz group, respectively.

Rewriting the relevant operators in spinor notation one obtains ($f\to g f$)
\begin{align}
F_B(\mu) \Phi_+(z;\mu) & = i\,\langle 0| \bar\psi_+(z) h_+(0) - \chi_+(z) \bar h_+(0)|\bar B(v)\rangle\,,
\notag\\[2mm]
F_B(\mu) \Phi_-(z;\mu) & = i\,\langle 0| \bar\psi_-(z) h_-(0) - \chi_-(z) \bar h_-(0)|\bar B(v)\rangle\,,
\notag\\[2mm]
   {2} F_B(\mu) \Phi_3(z_1,z_2;\mu) &=-
\langle 0|\chi_+(z_1)\bar f_{++}(z_2)  h_+(0) + \bar\psi_+(z_1) f_{++}(z_2) \bar h_+(0) |\bar B(v)\rangle\,,
\label{spinor23}
\end{align}
and
\begin{align}
    2 F_B(\mu) \Phi_4(z_1,z_2;\mu) &=
\phantom{-}  \langle 0|\chi_-(z_1)  f_{++}(z_2) h_-(0) + \bar\psi_-(z_1) \bar f_{++}(z_2) \bar h_-(0) |\bar B(v)\rangle\,,
\notag\\[2mm]
   F_B(\mu) \big[ \Psi_4 + \widetilde{\Psi}_4\big](z_1,z_2;\mu) &=
    - \langle 0| \chi_+(z_1)f_{+-}(z_2) h_-(0) +  \bar\psi_+(z_1)\bar f_{+-}(z_2) \bar h_-(0) |\bar B(v)\rangle\,,
\notag\\[2mm]
   F_B(\mu) \big[ \Psi_4 - \widetilde{\Psi}_4\big](z_1,z_2;\mu) &=
    - \langle 0| \chi_+(z_1) \bar f_{+-}(z_2)h_-(0) + \bar\psi_+(z_1) f_{+-}(z_2) \bar h_-(0)|\bar B(v)\rangle\,.
\label{spinor4}
\end{align}
Since the contributions of left- and right-handed (chiral and anti-chiral)
quarks have to be equal, one can drop half of the terms, e.g., the ones involving
$\psi$-spinor, for most purposes.
Note that $\Phi_4$ and $\Psi_4 + \widetilde{\Psi}_4$ contain light quark and gluon fields of the same chirality,
whereas in $\Phi_3$ and $\Psi_4 -  \widetilde{\Psi}_4$ chirality of the light degrees of freedom is the opposite.
Since chirality (and twist) is conserved in perturbation theory, we expect that
$\Phi_4$ can get mixed under evolution with $\Psi_4 + \widetilde{\Psi}_4$, but
the scale dependence of the ``genuine'' twist-four contribution to $\Psi_4 - \widetilde{\Psi}_4$ is autonomous.

For completeness we write also twist-five and twist-six DAs in the spinor representation:
\begin{align}
   2 F_B(\mu) \Phi_5(z_1,z_2;\mu) &=
  \langle 0|\chi_+(z_1)  f_{--}(z_2) h_+(0) + \bar\psi_+(z_1)\bar  f_{--}(z_2) \bar h_+(0) |\bar B(v)\rangle\,,
\notag\\[2mm]
   F_B(\mu) \big[ \Psi_5 + \widetilde{\Psi}_5\big](z_1,z_2;\mu) &=
   \langle 0| \chi_-(z_1)f_{+-}(z_2)h_+(0) + \bar\psi_-(z_1)\bar f_{+-}(z_2) \bar h_+(0) |\bar B(v)\rangle\,,
\notag\\[2mm]
   F_B(\mu) \big[ \Psi_5 - \widetilde{\Psi}_5\big](z_1,z_2;\mu) &=
    \langle 0| \chi_-(z_1) \bar f_{+-}(z_2)h_+(0) + \bar\psi_-(z_1) f_{+-}(z_2) \bar h_+(0)|\bar B(v)\rangle\,,
\label{spinor5}
\end{align}
and
\begin{align}
{2} F_B(\mu) \Phi_6(z_1,z_2;\mu) &=~
\langle 0| \chi_-(z_1)\bar f_{--}(z_2)  h_-(0) + \bar\psi_-(z_1) f_{--}(z_2) \bar h_-(0) |\bar B(v)\rangle\,.
\label{spinor6}
\end{align}
%
\section{Scale Dependence}\label{sec:scale}
%
DAs are scale-dependent and satisfy renormalization group equations with the evolution kernels 
that can be found in Refs.~\cite{Braun:2009vc, Knodlseder:2011gc} (in position space).
They are collected in Appendix~\ref{app:kernels}.
The corresponding expressions in momentum space can be found in Ref.~\cite{Ji:2014eta}.

The evolution equation for the leading-twist B-meson DA $\Phi_+$ was derived by Lange and Neubert~\cite{Lange:2003ff}
and solved in Refs.~\cite{Bell:2013tfa,Braun:2014owa}. The evolution equation for the twist-three DAs $\Phi_{-}$ and $\Phi_3$
was constructed and solved in the large-$N_c$ limit in Ref.~\cite{Braun:2015pha} using a ``hidden'' symmetry of this
equation called complete integrability. It turns out that evolution equations for the twist-four DAs are completely
integrable as well and can be solved in the same manner. The corresponding expressions present the main result of
our study. In this section we present the final expressions.
The derivation uses the formalism of the Quantum Inverse Scattering Method (QISM) and is sketched
in Appendix~\ref{app:RGequations};  details will be published elsewhere.

For the two-particle twist-two and twist-three DAs one obtains
\cite{Beneke:2000wa,Bell:2013tfa,Braun:2014owa,Braun:2015pha}
\begin{align}
  \Phi_+(z,\mu) &= -\frac{1}{z^2}\int_0^\infty ds\, s\,e^{is/z}\, \eta_+(s,\mu)\,,
\notag\\
  \Phi_-(z,\mu) &= -\frac{i}{z}\int_0^\infty ds\,  e^{is/z}\, \big[\eta_+(s,\mu)+ \eta_3^{(0)}(s,\mu)\big],
\label{Phi+Phi-}
\end{align}
in position space, and
\begin{align}
  \phi_+(\omega,\mu) &= \int_0^\infty ds\,\sqrt{\omega s} J_1(2\sqrt{\omega s})\, \eta_+(s,\mu)\,,
\notag\\
  \phi_-(\omega,\mu) &= \int_0^\infty ds\,J_0(2\sqrt{\omega s})\, \big[\eta_+(s,\mu)+ \eta_3^{(0)}(s,\mu)\big]
\notag\\&=
 \int_\omega^\infty \frac{d\omega'}{\omega'} \phi_+(\omega',\mu) + \int_0^\infty ds\,J_0(2\sqrt{\omega s})\,\eta_3^{(0)}(s,\mu)
\label{phi+phi-}
\end{align}
in momentum space, respectively. 
The coefficient functions $\eta_+(s,\mu)$ and $\eta_3^{(0)}(s,\mu)$ contain all relevant nonperturbative
information and have to be fixed at a certain (low) reference scale $\mu=\mu_0$.%
\footnote{The variable $s$ is an eigenvalue of the generator of special conformal transformation $S_+$~\cite{Braun:2014owa}.
The leading-twist coefficient function $\eta_+$ is related to the B-meson DA in
the ``dual'' representation of Ref.~\cite{Bell:2013tfa} as $s\, \eta_+(s,\mu) \equiv \rho_+(1/s, \mu)$.}
The most important parameter for the QCD description of $B$-decays is the value of the first negative moment
\begin{align}
  \lambda_B^{-1}(\mu) &= \int_0^\infty \frac{d\omega}{\omega} \phi_+(\omega,\mu) = \int_0^\infty d\tau \, \Phi_+(-i\tau, \mu) =
  \int_0^\infty ds\, \eta_+(s,\mu)\,.
\end{align}

The scale dependence of  $\eta_+(s,\mu)$ and $\eta_3^{(0)}(s,\mu)$  is given by
\begin{align}
   \eta_+(s,\mu) &= R(s;\mu,\mu_0)\eta_+(s,\mu_0) \,,
\notag\\
   \eta_3^{(0)}(s,\mu) &= L^{N_c/\beta_0}R(s;\mu,\mu_0)\eta_3^{(0)}(s,\mu_0)\,,
\end{align}
where $L = {\alpha_s(\mu)}/{\alpha_s(\mu_0)}$ and
\begin{eqnarray}
R(s;\mu,\mu_0) &=& {L^{3C_F/(2\beta_0)}}  \exp\left[-\int_{\mu_0}^{\mu} \frac{d\tau}{\tau}\,
\Gamma_{cusp}(\alpha_s(\tau))\,\ln (\tau s/s_0) \right]
\nonumber\\&=& {L^{3C_F/(2\beta_0)}} \left(\frac{\mu}{\mu_0}\right)^{-\frac{2C_F}{\beta_0}}
\left(\frac{\mu_0 s}{s_0}\right)^{\frac{2C_F}{\beta_0}\ln L}\!\!
L^{-\frac{4C_F\pi}{\beta_0^2\,\alpha_s(\mu_0)}}\,.
\label{Rfactor}
\end{eqnarray}
Here
${s_0= e^{5/4-\gamma_E}}$, $\Gamma_{cusp}(\alpha_s)=\frac{\alpha_s}{\pi} C_F+\ldots $
is the cusp anomalous dimension
and we have factored out the scale dependence of the B-meson decay constant
\begin{align}
 F_B(\mu) &=  L^{-3C_F/(2\beta_0)} F_B(\mu_0) \,.
\end{align}

The three-particle twist-three DA $\Phi_3(z_1, z_2,\mu)$ satisfies a more complicated renormalization group (RG) equation
\begin{align}
 \Big(\mu\frac{\partial}{\partial\mu}+\beta(\alpha_s)\frac{\partial}{\partial \alpha_s}
+\frac{\alpha_s}{2\pi} \mathcal{H}_3\Big)\Phi_3(\underline{z},\mu) =0\,, \qquad  \underline{z} = \{z_1,z_2\}\,,
\end{align}
where ${\cal H}_3$ is a certain integral operator that can be written as a sum of two-particle kernels that can be found in Appendix~\ref{app:kernels}.
This equation was solved in Ref.~\cite{Braun:2015pha} in the large-$N_c$ limit, i.e. neglecting corrections to ${\cal H}_3$
that are suppressed by a factor $1/N_c^2$. In~\cite{Braun:2015pha} the eigenfunctions of the evolution equation
and the corresponding anomalous dimensions have been found.

The DA $\Phi_3(z_1, z_2,\mu)$ can be expanded in terms the eigenfunctions of the
large-$N_c$ evolution  kernel as follows~\cite{Braun:2015pha}:
\begin{eqnarray}
\hspace*{-5mm}\Phi_3(\underline{z},\mu)
&=&
\int_0^\infty\!\! ds \Big[ \eta_3^{(0)}(s,\mu)\,Y_3^{(0)}(s\,|\,\underline{z})
+ \frac12 \int_{-\infty}^\infty\!\! dx\,\eta_3(s,x,\mu)\,Y_{3}(s,x\,|\,\underline{z})\Big],
\label{Phi3}
\end{eqnarray}
where
\begin{align}
 Y_{3}(s,x\,|\,\underline{z}) &=\frac{is^2}{z_1^2 z_2^3}\int_0^1 du\,u\bar  u\, e^{is(u/z_1+\bar u/z_2)}\,
{}_2F_1\left(\genfrac{}{}{0pt}{}{-\frac12-ix,-\frac12+ix}{2}\Big|-\frac{u}{\bar u}\right),
\qquad \bar u\equiv1-u\,,
\notag\\
Y^{(0)}_3(s\,|\,\underline{z})&=  Y_{3}(s,x={i}/{2}\,|\,\underline{z}) = \frac{is^2}{z_1^2 z_2^3}\int_0^1 du\, u\bar  u\,e^{is(u/z_1+\bar u/z_2)}\,.
\end{align}
Note that the eigenfunctions $Y_{3}(s,x\,|\,\underline{z})$ are even under reflection $x\to -x$, so that the 
coefficient functions in this expansion can be chosen even as well, $\eta_3(s,x,\mu) = \eta_3(s,-x,\mu)$.
They are characterized by two real numbers $s>0$ and $-\infty < x < \infty$.
It turns out that the corresponding anomalous dimensions can be written as a sum of terms depending on $s$ and $x$ separately.
The $s$-dependent part can be absorbed in the same universal factor $R(s;\mu,\mu_0)$ as for the leading twist so
that one obtains~\cite{Braun:2015pha}
\begin{align}
 \eta_3(s,x,\mu) &= L^{\gamma_3(x)/\beta_0}R(s;\mu,\mu_0)\,\eta_3(s,x,\mu_0)\,,
\notag\\
   \eta_3^{(0)}(s,\mu) &= L^{N_c/\beta_0}R(s;\mu,\mu_0)\eta_3^{(0)}(s,\mu_0)\,,
\label{scale:twist23}
\end{align}
with the anomalous dimension~\cite{Braun:2015pha}%
\footnote{To be precise, $\gamma_3(x)$ and $\gamma_4(x)$ defined below in Eq.~\eqref{gamma4} 
 correspond to the \emph{difference} of anomalous
 dimensions between the  higher- and leading-twist operators. The scale dependence for the leading-twist DA
 is included in the $R$-factor.}
\begin{align}
 \gamma_3(x) &=  N_c\big[\psi\big(3/2+ix\big)+ \psi\big(3/2-ix\big) +2\gamma_E\big]\,,&&
\gamma_3^{(0)}=\gamma_3(x= {i}/{2}) = N_c\,.
\label{gamma3}
\end{align}
Note that in addition to the integral over all real values of $x$ the DA $\Phi_3(\underline{z},\mu)$ contains
an extra contribution, the first term in Eq.~\eqref{Phi3}, corresponding to a particular imaginary value $x=i/2$.
This special term has a lower anomalous dimension separated by a finite number from the rest, continuum spectrum,
and  can be interpreted as the asymptotic DA. This interpretation fails, however, for large quark and/or gluon momenta $\omega_1,\omega_2 \gtrsim \mu$
(alias small coordinates $z_1,z_2 \lesssim 1/\mu$) in which case contributions with all anomalous dimensions have to be included,
see Ref.~\cite{Braun:2015pha} for a detailed discussion. Note also that the
twist-three contribution to the two-particle DA $\Phi_-$ in Eq.~\eqref{Phi+Phi-} is determined entirely by this special term, $\eta_3^{(0)}(s,\mu)$.
The  ``genuine'' three-particle twist-three contributions to $\Phi_3$ encoded in $\eta_3(s,x,\mu)$
decouple from $\Phi_-$ to the stated $1/N_c^2$ accuracy.

\begin{figure}[t]
\centerline{
\begin{picture}(210,140)(0,0)
\put(-5,0){\epsfxsize7.8cm\epsffile{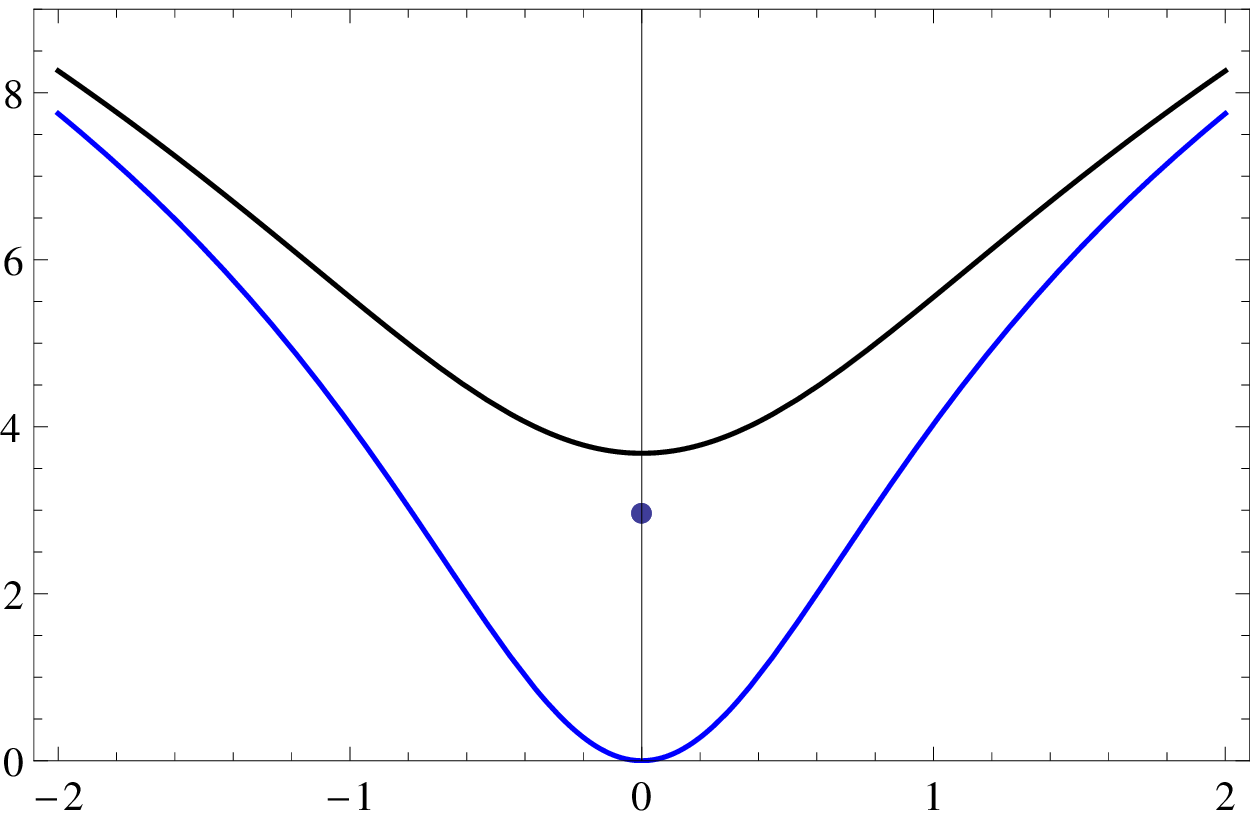}}
\put(105,-8){$x$}
\put(55,100){$\gamma_3(x)$}
\put(30,60){\textcolor{blue}{$\gamma_4(x)$}}
\end{picture}
}
\caption{
Anomalous dimensions $\gamma_3(x)$ \eqref{gamma3} and $\gamma_4(x)$ \eqref{gamma4} of higher-twist DAs.
The anomalous dimension of the special (discrete) twist-three state $\gamma_3^{(0)} = \gamma_3(x=i/2)$ is shown by the 
black dot.   
}
\label{fig:anom-dim}
\end{figure}

It is useful to have in mind that the evolution kernel $\mathcal{H}_3$ is a hermitian operator
so that its eigenfunctions $Y_{3}(s,x\,|\,\underline{z})$ are mutually orthogonal and form a complete set of functions
with respect to a suitable scalar product. Explicit construction is given in Appendix~\ref{app:scalarproduct}.
In this way the coefficient functions  $\eta_3(s,x,\mu)$, $\eta_3^{(0)}(s,\mu)$ can be calculated as scalar
products of the model DA with the corresponding eigenfunctions.

The renormalization group equations for twist-four DAs can be treated along the similar lines, the main difference 
being that one obtains a $2\times2$ matrix RG equation. The matrix structure is clear for the chiral-even case
since the two existing chirality-even DAs $\Phi_4$ and $\Psi_4 + \widetilde{\Psi}_4$ are mixed by the evolution,
and also for the chiral-odd case it is necessary to take into account additional operators containing 
transverse derivatives \eqref{def:Xi4}, see App.~\ref{app:RGequations}. The complete solution of the evolution equations 
is presented in Eq.~\eqref{eq:t4general}.  

It turns out that this general result can be simplified under the assumption that twist-four four-particle B-meson DAs 
(with two gluon fields and/or with an extra quark-antiquark pair) are put to zero. Such DAs can be expected to be small
and have autonomous scale dependence i.e. they do not mix with the three-particle DAs~\cite{Bukhvostov:1985rn}. 
Thus they can be consistently put to zero at 
all scales and do not reappear via evolution.  

With this assumption we are able to derive an exact relation between the two chiral-even DAs,
\begin{align}
2\partial_1 z_1 \Phi_4(\underline{z}) &= \Big(z_2\partial_{z_2}+2\Big)\left[\Psi_4 (\underline{z})+\widetilde\Psi_4 (\underline{z})\right],
\label{PhiPsirelation}
\end{align}
and a similar relation for the chiral-odd DAs that allows one to eliminate contributions with transverse derivatives~\eqref{WWidentities}.
The derivation is presented in App.~\ref{app:WW}.

In this way (to this accuracy) the general expressions for the scale dependence of the DAs in \eqref{eq:t4general} reduce to
\begin{subequations}
\label{eq:t4final}
\begin{align}
   \Phi_4(\underline{z})
&=
\frac12 \int\limits^\infty_0{ds}\int\limits^\infty_{-\infty}dx\,
\eta_4^{(+)}(s,x,\mu)\,{Y}_{4;1}^{(+)}(s,x\,|\underline{z})\,,
\label{Phi4final}\\
 (\Psi_4 + \widetilde{\Psi}_4)(\underline{z})&=
- \int\limits^\infty_0{ds}\int\limits^\infty_{-\infty}dx\,
  \eta_4^{(+)}(s,x,\mu)\,{Y}_{4;2}^{(+)}(s,x\,|\underline{z})\,,
\label{Psi+tPsifinal}\\
 (\Psi_4 - \widetilde{\Psi}_4)(\underline{z})&=
 2 \int\limits^\infty_0{ds}\left(\frac{iz_2}{s}\right)
\biggl\{ \eta_3^{(0)}(s,\mu)\,{Y}_{3}^{(0)}(s\,|\underline{z}) + \frac12\int\limits^\infty_{-\infty}dx\,
\eta_3(s,x,\mu)\,{Y}_{3}(s,x\,|\underline{z})\biggr\}
\notag\\&\quad
 - \int\limits^\infty_0{ds}\int\limits^\infty_{-\infty}dx\, \varkappa_4^{(-)}(s,x,\mu)\, {Z}_{4;2}^{(-)}(s,x\,|\underline{z})\,,
\label{Psi-tPsifinal}
\end{align}
\end{subequations}
where
\begin{align}
\label{eq:Y+states}
\begin{pmatrix} Y_{4;1}^{(+)}\\ Y_{4;2}^{(+)}\end{pmatrix}(s,x\,|\,\underline{z}) &=
\frac{s^{3/2}}{z_1^2z_2^3}\int^1_0du\,\e^{is(u/z_1+\bar u/z_2)}\,
{}_2F_1\left(\genfrac{}{}{0pt}{}{ -ix, +ix}{1}\Big|-\frac{u}{\bar u}\right)
\begin{pmatrix}\bar u z_1 \\[1mm] -u z_2\end{pmatrix},
\end{align}
and
\begin{align}
\label{eq:Z-states}
Z_{4;2}^{(-)} (s,x\,|\,\underline{z}) &=
\frac{s^{3/2}}{z_1^2z_2^2}\int^1_0du\, {u^2}\,\e^{is(u/z_1+\bar u/z_2)}\,
{}_2F_1\left(\genfrac{}{}{0pt}{}{-ix,+ix}{3}\Big|-\frac{u}{\bar u}\right).
\end{align}
The first term in the expression for  $\Psi_4 - \widetilde{\Psi}_4$ \eqref{Psi-tPsifinal} 
can be interpreted as the Wandzura-Wilczek-type contribution of twist-three (related to the DA $\Phi_3$) 
in the same manner as the twist-three two-particle DA $\Phi_-$ contains a term related to the
leading-twist DA $\Phi_+$, cf. Eq.~\eqref{Phi+Phi-}. We can write
\begin{align}
  (\Psi_4 - \widetilde{\Psi}_4)(\underline{z})&= (\Psi_4 - \widetilde{\Psi}_4)_{WW}(\underline{z}) + (\Psi_4 - \widetilde{\Psi}_4)_{tw-4}(\underline{z})\,.
\label{WWdecomposition}
\end{align}

The remaining ``genuine'' twist-four contributions are expressed in terms of two 
nonperturbative functions $\eta_4^{(+)}(s,x,\mu)$ and $\varkappa_4^{(-)}(s,x,\mu)$ that 
have the same scale dependence
(up to $1/N_c^2$ corrections):
\begin{align}
   \eta_4^{(+)} (s,x,\mu) &= L^{\gamma_4(x)/\beta_0}R(s;\mu,\mu_0)\,  \eta_4^{(+)}(s,x,\mu_0)\,,
\notag\\
   \varkappa_4^{(-)} (s,x,\mu) &= L^{\gamma_4(x)/\beta_0}R(s;\mu,\mu_0)\,  \varkappa_4^{(-)}(s,x,\mu_0)\,,
\label{scale:twist4}
\end{align}
where 
\begin{align}
 \gamma_4(x) &=  N_c\big[\psi\big(ix\big)+ \psi\big(-ix\big) +2\gamma_E\big]\,.
\label{gamma4}
\end{align}
Since the scale dependence is the same, one may wonder whether a relation 
between $\eta_4^{(+)}(s,x,\mu)$ and $\varkappa_4^{(-)}(s,x,\mu)$ exists on a nonperturbative level. 
Such relation is derived and discussed in the next section; its theoretical status is, however, less
solid as compared to the rest of our results. 

Going over to momentum space corresponds to a Fourier transform of the eigenfunctions
\begin{align}
   Y(s,x\,|\,\underline{z}) = \int_0^\infty \!\!d\omega_1\!  \int_0^\infty \!\!d\omega_2\,\, e^{-i\omega_1 z_1-i\omega_2 z_2}\,
   {Y} (s,x\,|\,\underline{\omega})\,, \qquad \underline{\omega}=\{\omega_1,\omega_2\}\,,
\end{align}
so that, e.g.,
\begin{eqnarray}
\hspace*{-5mm}\phi_3(\underline{\omega},\mu)
&=&
\int_0^\infty\!\! ds \Big[ \eta_3^{(0)}(s,\mu)\,{Y}_3^{(0)}(s\,|\,\underline{\omega})
+ \frac12 \int_{-\infty}^\infty\!\! dx\,\eta_3(s,x,\mu)\,{Y}_{3}(s,x\,|\,\underline{\omega})\Big],
\label{Phi3mom}
\end{eqnarray}
and similar for all twist-four DAs.%
\footnote{The factor $iz_2$ in the expression for $(\Psi_4 - \widetilde{\Psi}_4)(\underline{z})$
has to be replaced in the momentum space representation
by the derivative over the gluon momentum $iz_2\mapsto -\partial/\partial\omega_2$.}

Explicit expressions for the eigenfunctions of the evolution equations in momentum space can easily be derived
using that
\begin{align}
 \frac{e^{-i\pi j}}{z^{2j}} e^{is/z} = \int_0^\infty \!\!d\omega\, e^{-i\omega z}\,(\omega/s)^{j-1/2}\, J_{2j-1}(2\sqrt{s\omega})
\qquad \text{Im}(z)<0\,.
\end{align}
One obtains
\begin{align}
{Y}_{3}(s,x\,|\,\underline{\omega}) &=
 -\!\int\limits_0^1\!\! du\,\sqrt{us\omega_1}\,J_1(2\sqrt{us\omega_1})\,\omega_2\,J_2(2\sqrt{\bar u s \omega_2})\,
{}_2F_1\left(\genfrac{}{}{0pt}{}{-\frac12\!-\!ix,-\frac12\!+\!ix}{2}\Big|-\frac{u}{\bar u}\right)\!,
\notag\\
\begin{pmatrix} {Y}_{4;1}^{(+)}\\ {Y}_{4;2}^{(+)}\end{pmatrix}\!(s,x\,|\,\underline{\omega}) &=
\!\int\limits^1_0\frac{du}{\bar u}\sqrt{\bar u s\,\omega_2}\,
{}_2F_1\left(\genfrac{}{}{0pt}{}{\!-ix,+ix}{1}\Big|-\frac{u}{\bar u}\right)\!
\begin{pmatrix}
    \sqrt{\omega_2\bar u}J_0\left(2\sqrt{\omega_1us}\right)J_2\left(2\sqrt{\omega_2\bar us}\right)\\[2mm]
   -\sqrt{\omega_1u}J_1\left(2\sqrt{\omega_1us}\right)J_1\left(2\sqrt{\omega_2\bar us}\right)
\end{pmatrix}\!,
\notag\\
{Z}_{4;2}^{(-)} (s,x\,|\,\underline{\omega}) &=
\int\limits^1_0\frac{du}{\bar u}\,u\, \sqrt{u s \omega_1}J_1\left(2\sqrt{\omega_1us}\right)\,
\sqrt{\bar u \omega_2}\,J_1\left(2\sqrt{\omega_2\bar us}\right)
{}_2F_1\left(\genfrac{}{}{0pt}{}{-ix,+ix}{3}\Big|-\frac{u}{\bar u}\right).
\end{align}
The RG kernels (for all twists)  are hermitian operators with respect to the $SL(2)$ scalar product, see
App.~\ref{app:RGequations}, \ref{app:scalarproduct}. As the result, eigenfunctions of the evolution equation corresponding to 
different anomalous dimensions are orthogonal and form a complete set of functions.
The resulting orthogonality relations are collected in  App.~\ref{app:scalarproduct}.
They can be used to invert the representations in Eq.~\eqref{eq:t4final} and express the
coefficients $\eta_4$, $\varkappa_4$ at a low reference scale in terms of the models for the DAs in 
momentum (or coordinate) space.

%
\section{Two-particle higher-twist DAs and equations of motion}\label{sec:KKQT}
%

In the parton model language, higher twist effects are due to contributions of higher Fock
states but also to nonvanishing parton transverse momenta (or virtuality). Due to QCD equations
of motion (EOM) the latter can be expressed in terms of the multiparton configurations as well and
can be thought of as contributions of gluon (or quark-antiquark pair) emission from the external
lines of the partonic hard-scattering amplitude. In applications to hard exclusive reactions it
has become customary to take into account these diagrams through the contributions of two-particle higher-twist DAs
that arise as terms $\sim \mathcal{O}(x^2)$ in the expansion of the relevant
nonlocal quark-antiquark operator close to the light cone. For the present case we can write,
assuming $|x^2| \ll 1/\Lambda_{\rm QCD}^2$,
\begin{eqnarray}
 \langle 0| \bar q(x) \Gamma [x,0] h_v(0) |\bar B(v)\rangle &=&
-\frac{i}2  F_B \Tr\Big[\gamma_5 \Gamma P_+ \Big] \int\limits_0^\infty d\omega \, e^{-i\omega (vx)}
\Big\{\phi_+(\omega) + x^2 g_+(\omega)\Big\}
\nonumber\\&&{}\hspace*{-4cm} +\frac{i}4  F_B \Tr\Big[\gamma_5 \Gamma P_+ \slashed{x} \Big] \frac{1}{vx}
\int\limits_0^\infty d\omega \, e^{-i\omega (vx)} \Big\{[\phi_+-\phi_-](\omega) + x^2 [g_+-g_-](\omega)\Big\}
\label{def:g+g-}
\end{eqnarray}
introducing two new DAs, $g_+(\omega)$ and $g_-(\omega)$ that are of twist four and five, respectively.
Terms $\mathcal{O}(x^4)$ are neglected.
Eq.~\eqref{def:g+g-} has to be understood as a light-cone expansion to the tree-level accuracy
which should be sufficient for the calculation of higher-twist corrections to the leading order in the
strong coupling.

Note that the l.h.s. of Eq.~\eqref{def:g+g-} cannot have a power singularity $1/(vx)$ at $(vx)\to 0$.
This implies the constraints
\begin{align}
   \int\limits_0^\infty d\omega \,  \Big[\phi_+(\omega) -\phi_-(\omega)\Big] = 0\,,
&&
   \int\limits_0^\infty d\omega \,  \Big[g_+(\omega) -g_-(\omega)\Big] = 0\,.
\end{align}

The DAs  $g_+(\omega)$ and $g_-(\omega)$ can be expressed in terms of the three-particle DAs considered in
previous sections. The corresponding expressions were derived by Kawamura \emph{et al.}~(KKQT)~\cite{Kawamura:2001jm}
starting from the operator identities
\begin{align}
  \frac{\partial}{\partial x^\mu} \bar q(x)\gamma^\mu \Gamma [x,0] h_v(0)
&=
- i\int_0^1\! udu\, \bar q(x)[x,ux] x^\rho gG_{\rho\mu}(ux)[ux,0] \gamma^\mu \Gamma h_v(0)\,,
\notag\\
 v^\mu \frac{\partial}{\partial x^\mu} \bar q(x) \Gamma [x,0] h_v(0)
&=
 \phantom{-} i\int_0^1\! \bar udu\, \bar q(x)[x,ux] x^\rho gG_{\rho\mu}(ux)[ux,0] v^\mu \Gamma h_v(0)
\notag\\& \qquad{}
 + (v\cdot\partial) \bar q(x) \Gamma [x,0] h_v(0)\,,
\label{identity}
\end{align}
taking the appropriate matrix elements and comparing the resulting expressions with the definition of the
DAs in the limit $x^2\to 0$.
In this way one obtains~\cite{Kawamura:2001jm}%
\footnote{The last two relations in~\eqref{KKQT} follow from the expressions given in~\cite{Kawamura:2001jm} by simple algebra.}
\begin{subequations}
\label{KKQT}
\begin{align}
\hspace*{-0.5cm}  \Big[z\frac{d}{dz}+1\Big]\Phi_-(z) &=  \Phi_+(z)  + 2 z^2  \int_0^1\! udu\,\Phi_3(z,uz)\,,
\label{KKQT1}
\\
2 z^2  \mathrm{G}_+(z) & =
-  \Big[ z \frac{d}{dz} - \frac12  + i z \bar \Lambda \Big] \Phi_+(z)
-  \frac{1}{2}\Phi_-(z)
- z^2  \int_0^1\! \bar udu\,{\Psi}_4(z,uz)\,,
\label{KKQT2}
\\
 2 z^2 \mathrm{G}_-(z)
&= -  \Big[ z \frac{d}{dz} - \frac12  + i z \bar \Lambda \Big] \Phi_-(z) - \frac12  \Phi_+(z)
- z^2  \int_0^1\! \bar udu\,{\Psi}_5(z,uz)\,,
\label{KKQT3}
\\
 \Phi_-(z)
&= \left(z \frac{d}{dz}+1 + 2i z \bar \Lambda  \right) \Phi_+(z) +
2 z^2 \int_0^1\! du\,  \Big[ u \Phi_4(z,uz) + {\Psi}_4(z,uz)\Big],
\label{KKQT4}
\end{align}
\end{subequations}
where
\begin{align}
  \mathrm{G}_\pm(z,\mu) &= \int\limits_0^\infty d\omega \, e^{-i\omega z}g_\pm(\omega,\mu)
\end{align}
and
\begin{align}
   \bar\Lambda = m_B -m_b\,.
\end{align}
%
The first KKQT relation, Eq.~\eqref{KKQT1}, only involves twist-two and twist-three contributions.
It allows to calculate the twist-three DA $\Phi_-(z)$ in terms of $\Phi_+(z)$ (Wandzura-Wilczek
contribution~\cite{Beneke:2000wa}) and the ``genuine'' twist-three three-particle DA $\Phi_3(z_1,z_2)$.
This relation can be derived in many ways (see e.g. App.~\ref{app:WW}) and was used to arrive at the 
representation for $\Phi_-(z)$ in Eqs.~\eqref{Phi+Phi-},~\eqref{phi+phi-}~\cite{Braun:2015pha}.

The second and the third relation,  Eqs.~\eqref{KKQT2} and \eqref{KKQT3}, provide one with the expressions
for the two-particle higher-twist DAs $\mathrm{G}_\pm(z)$ in terms of the three-particle DAs of the same twist
and lower-twist Wandzura-Wilczek-type terms.

The last KKQT relation,  Eq.~\eqref{KKQT4}, is a nontrivial constraint relating  the
higher-twist matrix elements with the leading twist.
Using the representation in Eqs.~\eqref{Phi+Phi-},~\eqref{eq:t4final} one obtains from  Eq.~\eqref{KKQT4} the
following relation for the coefficient functions:%
\footnote{In this calculation one has to start with the regularized version of the integral
$ \int_0^1\! du\,{\Psi}_4(z,uz)\to \int_0^1\! du\,u^\epsilon\,{\Psi}_4(z,uz) $ and take the limit $\epsilon \to 0$
at the end.}
\begin{align}
\left[1 - (\partial_s s)^2 - 2 s \bar\Lambda \right]\,   \eta_+(s,\mu) 
&= \pi  \sqrt{s} \varkappa_4^{(-)}(s,0,\mu) - \pi \sqrt{s}\eta_4^{(+)}(s,0,\mu)\,.
\label{secondEOM}
\end{align}
This equation presents a nonlocal generalization of the Grozin-Neubert relations~\cite{Grozin:1996pq}
\begin{align}
  \int_0^\infty\! d\omega\, \omega\,  \phi_+(\omega) = \frac43\bar\Lambda\,, \qquad
  \int_0^\infty\! d\omega\, \omega^2 \phi_+(\omega) = 2\bar \Lambda^2 + \frac23\lambda_E^2 + \frac13 \lambda_H^2\,,
\label{GN}
\end{align}
where $\lambda_E^2$ and $\lambda_H^2$ are matrix elements of certain local quark-gluon operators \eqref{def:lambdaEH}. 
It is easy to show that Eq.~\eqref{GN} correspond to the expansion of Eq.~\eqref{secondEOM} at $s\to 0$ and collecting terms $\mathcal{O}(s)$ and
 $\mathcal{O}(s^2)$, respectively.

Since $\gamma_4(x=0)=0$ \eqref{gamma4}, the scale dependence of the higher-twist contributions on the r.h.s. of Eq.~\eqref{secondEOM} matches the scale
dependence of the leading-twist DA on the l.h.s. of this relation, however, only if the derivatives $\partial_s$ are not applied to the 
$s$-dependent $R$-factor \eqref{Rfactor}. This difficulty is due to the fact the light-cone expansion $x^2\to 0$ in \eqref{def:g+g-} 
beyond tree level requires a careful treatment of the $x^2$-dependent cusp anomalous dimension.   
A detailed investigation of this problem goes beyond the tasks of this work.

In order to tame potentially large corrections $\sim \alpha_s\ln (s\mu)$ to Eq.~\eqref{secondEOM} one
can try to enforce this relation for the integrated quantities, $\int_0^\infty ds$, in which case
it transforms into a constraint on the low-momentum behavior of the DAs that is most
relevant for applications. In this way one obtains after a short calculation
\begin{align}
 2\bar\Lambda \phi_+'(0,\mu) - \lambda_B^{-1}(\mu) &=
  2 \int_0^\infty \frac{d\omega_2}{\omega_2^2} \Big[\phi_4(0,\omega_2,\mu) + 
  \omega_2 (\psi_4)_{\rm{tw-4}}^{(1)}(0,\omega_2,\mu)\Big]
\label{secondEOM1}
\end{align}
 where $\phi'(\omega,\mu) = \partial_{\omega}\phi(\omega,\mu)$ and
 $(\psi_4)_{\rm{tw-4}}$ is the ``genuine'' twist-four contribution to the DA $\psi_4$,
$$(\psi_4)_{\rm{tw-4}}^{(1)}(\underline{\omega},\mu) = \partial_{\omega_1}(\psi_4)_{\rm{tw-4}}(\underline{\omega},\mu)\,,
\qquad 
(\psi_4)_{\rm{tw-4}} = \frac12 \left[(\psi_4\!-\!\widetilde\psi_4)_{\rm{tw-4}}+ (\psi_4\!+\!\widetilde\psi_4)\right],
$$
cf.~Eq.~\eqref{WWdecomposition}.

In any case, it is important to have in mind that the expressions for the two-particle higher-twist DAs $\mathrm{G}_\pm(z)$ in \eqref{KKQT2},\eqref{KKQT3}
are obtained in the same approximation as the constraints in \eqref{KKQT4}, \eqref{secondEOM}, or, in the minimal version, \eqref{secondEOM1}. These 
constraints have to be fulfilled, for consistency, for any model of $\mathrm{G}_\pm(z)$ at a low scale.    

In Ref.~\cite{Kawamura:2001jm} a model for the leading-twist DA was formulated, called there ``Wandzura-Wilczek approximation'',
by putting all quark-gluon contributions on the r.h.s. of Eqs.~\eqref{secondEOM} to zero.
We think that this approximation is not viable and the interpretation referring to Wandzura and Wilczek is misleading.

Similar EOM constraints are familiar and widely used for the light-quark systems,
see e.g.~\cite{Ball:1998ff,Braun:2004vf}. The simplest of them is the following operator identity for the divergence of the
quark energy momentum tensor (for massless quarks):
\begin{align}
  \partial^\mu \mathcal{O}_{\mu\nu} = 2 \bar q g G_{\mu\nu}\gamma^\mu q\,,
\label{example1}
\end{align}
where
\begin{align}
  \mathcal{O}_{\mu\nu} = \frac12 \bar q \gamma_\mu i \stackrel{\leftrightarrow}{D}_\nu q +
\frac12 \bar q \gamma_\nu i \stackrel{\leftrightarrow}{D}_\mu q\,.
\end{align}
Such identities exist for all leading twist operators. The general statement is that
the divergence (in mathematical sense) of a multiplicatively renormalizable leading twist
operator can be expressed as a sum of quark-gluon operators~\cite{Braun:2011dg}.
If all quark-gluon contributions are put to zero, one obtains an infinite series of conserved
currents. This is in fact the symmetry of a naive parton model (free quarks), and indeed sending $g\to0$ is the only way to get rid of
quark-gluon operators in a theoretically consistent way.  A free theory is, however, not a viable approximation for modeling of the bound states.
Note that the scale dependence must be neglected in this case as well. This approximation has been, therefore, never followed, 
to the best of our knowledge.
Instead, the EOM relations of this type have commonly been used as constraints that allow one to reduce the
number of independent twist-four matrix elements, see e.g.~\cite{Ball:1998ff,Braun:2004vf,Braun:2011dg}.
The situation with the heavy-light operators is analogous; Eq.~\eqref{secondEOM} provides one with powerful constraints 
on the higher-twist DAs by reducing the number of nonperturbative parameters but does not
imply any restrictions on the leading-twist DA itself.

%
\section{Models}\label{sec:Models}
%

Modeling higher-twist DAs requires certain nonperturbative input which is currently very limited.
Aim of this section is to present a few phenomenologically acceptable models that satisfy all known
constraints.

Matrix elements of local operators
\begin{eqnarray}
\lefteqn{\langle 0| \bar q(0) gG_{\mu\nu}(0)\Gamma h_v(0) |\bar B(v)\rangle =}
\nonumber\\&=&
-\frac{i}{6} F_B \lambda^2_H \Tr\Big[\gamma_5\Gamma P_+ \sigma_{\mu\nu}\Big]
-\frac{1}{6} F_B\Big( \lambda^2_H- \lambda^2_E\Big)
  \Tr\Big[\gamma_5\Gamma P_+(v_\mu\gamma_\nu-v_\nu\gamma_\mu)\Big]
\label{def:lambdaEH}
\end{eqnarray}
can be estimated from QCD sum rules. One obtains
\begin{align}
 \lambda^2_E = 0.11\pm 0.06~\text{GeV}^2, && \lambda^2_H =  0.18\pm 0.07~\text{GeV}^2,&& \text{\cite{Grozin:1996pq}}
\\
 \lambda^2_E = 0.03\pm 0.02~\text{GeV}^2, && \lambda^2_H = 0.06\pm 0.03~\text{GeV}^2, &&  \text{\cite{Nishikawa:2011qk}}
\end{align}
where in the second calculation some NLO corrections have been taken into account. Note that the ratio
\begin{align}
   R =  \lambda_E^2/ \lambda^2_H \simeq 0.5
\label{REH}
\end{align}
is almost the same in both cases and is generally more reliable than the values of the matrix elements themselves as many uncertainties cancel.

Assuming that the integrals over quark and gluon momenta at a low scale converge, one obtains normalization conditions for the DAs
\begin{align}
&\Psi_V(\underline{z}=0) = \int^\infty_0\!\!d\omega_1\int^\infty_0\!\!d\omega_2\,\psi_V(\underline{\omega})=\frac13\lambda_H^2\,,
\qquad\qquad 
\Psi_A(\underline{z}=0) =\frac13\lambda_E^2\,,
\notag\\
&X_A(\underline{z}=0) = Y_A(\underline{z}=0) = \widetilde X_A(\underline{z}=0) = 0\,,
\end{align}
or, equivalently,
\begin{align}
 &\Phi_3(\underline{z}=0) = \frac13 (\lambda_E^2-\lambda_H^2)\,, && \Phi_4(\underline{z}=0) = \frac13 (\lambda_E^2+\lambda_H^2)\,,
\notag\\
 &\Psi_4(\underline{z}=0) = \frac13 \lambda_E^2\,, &&  \widetilde\Psi_4(\underline{z}=0) = \frac13 \lambda_H^2\,. 
\end{align}
Under the same assumptions, at a low scale
\begin{align}
 G_+(0) = \int_0^\infty d\omega\, g_+(\omega) = \frac16 \Big[\bar\Lambda^2 + \lambda_E^2+\lambda_H^2\Big]\,.
\label{gplus-norm}
\end{align}

%
\subsection{Model I: Exponential}\label{subsec:Model-1}
%

The simplest model can be obtained combining the known low-momentum behavior \eqref{small-momenta} with an exponential falloff at large 
momenta, and using the above normalization conditions (cf.~\cite{Khodjamirian:2006st}):  
\begin{align}
\phi_3(\omega_1,\omega_2,\mu_0)&=\frac{\lambda_E^2-\lambda_H^2}{6\omega_0^5}\,\omega_1\omega_2^2\,\e^{-(\omega_1+\omega_2)/\omega_0}\,,
\notag\\
\phi_4(\omega_1,\omega_2,\mu_0)&=\frac{\lambda_E^2+\lambda_H^2}{6\omega_0^4}\,\omega_2^2\,\e^{-(\omega_1+\omega_2)/\omega_0}\,,
\notag\\
\psi_4(\omega_1,\omega_2,\mu_0)&=\frac{\lambda_E^2}{3\omega_0^4}\,\omega_1\omega_2\,\e^{-(\omega_1+\omega_2)/\omega_0}\,,
\notag\\
\widetilde\psi_4(\omega_1,\omega_2,\mu_0)&=\frac{\lambda_H^2}{3\omega_0^4}\,\omega_1\omega_2\,\e^{-(\omega_1+\omega_2)/\omega_0}\,,
\label{model1-momspace}
\end{align}
This construction is similar in spirit to the simple leading-twist DA proposed by Grozin and Neubert~\cite{Grozin:1996pq}
\begin{align}
\phi_+(\omega,\mu_0)&=\frac\omega{\lambda_B^2}\,\e^{-\omega/\lambda_B}\,, \qquad \eta_+(s,\mu)=\e^{-s\lambda_B}\,, 
\label{phi+GN}
\end{align} 
and has the advantage that all relevant coefficient functions in the expansion over the eigenfunctions of the evolution equations
can be calculated explicitly in analytic form:  
\begin{align}
\eta_3(s,x,\mu_0)&= - \frac{1}{18}(\lambda_E^2-\lambda_H^2) s^2 e^{-\omega_0 s} \frac{\pi (x^2+1/4)}{\cosh\pi x} \frac{x(x^2+9/4)}{\coth \pi x}\,, 
\notag\\
\eta^{(0)}_3(s,\mu_0)&=  - \frac{1}{18}(\lambda_E^2-\lambda_H^2) s^2 e^{-\omega_0 s}\,,
\notag\\
\eta_4^{(+)}(s,x,\mu_0)&=\frac13(\lambda_E^2+\lambda_H^2)s^{3/2}\e^{-\omega_0s}\coth \pi x \frac{\pi x^2}{\sinh\pi x}\,,
\notag\\
\varkappa_4^{(-)}(s,x,\mu_0)&={\frac1{36}(\lambda_H^2-\lambda_E^2)s^{3/2}\e^{-\omega_0s}
\frac{x(1+x^2)(4+x^2)}{\tanh \pi x}\frac{\pi x}{\sinh\pi x}}\,,
\label{model1-etaspace}
\end{align}
The EOM relation \eqref{secondEOM} becomes  
\begin{align}
\left[1 + \partial_s s  - \partial^2_s s^2 - 2 s \bar\Lambda \right]\,   \eta_+(s,\mu)
&=  \pi  \sqrt{s} \varkappa_4^{(-)}(s,0,\mu) - \pi \sqrt{s}\eta_4^{(+)}(s,0,\mu)
\notag\\& = -\frac29 (2\lambda_E^2+\lambda_H^2) s^{2}\e^{-\omega_0s}\,.
\label{mod1-EOM}
\end{align}
For the simplest leading-twist DA in \eqref{phi+GN} this equation is satisfied 
if
\begin{align}
 \omega_0 = \lambda_B = \frac23 \bar\Lambda\,, \qquad  2 \bar\Lambda^2 = 2\lambda_E^2+\lambda_H^2,
\label{impose}
\end{align}
in agreement with \eqref{GN}. If these relations are enforced, there remains to be one free parameter, e.g., the ratio 
$R = \lambda_E^2/\lambda_H^2$ \eqref{REH}.

\begin{figure}[t]
\centerline{
\begin{picture}(210,140)(0,0)
\put(-5,0){\epsfxsize7.8cm\epsffile{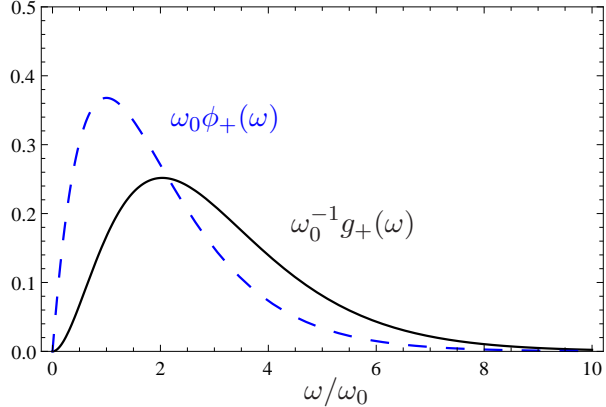}}
\put(105,-8){$\omega/\omega_0$}
\put(100,55){$\omega_0^{-1}g_+(\omega)$}
\put(55,95){\textcolor{blue}{$\omega_0\phi_+(\omega)$}}
\end{picture}
}
\caption{
The two-particle twist-4 DA $g_+(\omega)$ in Model I \eqref{gplus-model1}.
For this plot we have taken $R = \lambda_E^2/\lambda_H^2 = 1/2$ \eqref{REH}. 
The leading-twist DA $\phi_+(\omega)$  \eqref{phi+GN} is shown by dashes for comparison.
}
\label{fig:g+model1}
\end{figure}

Using the above explicit expressions for the three-particle DAs we can compute the two-particle twist-four DA $g_+(\omega)$ 
from Eq.~\eqref{KKQT2}.
Since this relation is derived in the same approximation as the EOM~\eqref{KKQT4}, we have to require that 
Eq.~\eqref{mod1-EOM} is satisfied identically. This means, e.g., that for the Grozin-Neubert leading-twist DA  
\eqref{phi+GN} the relations in \eqref{impose} have to be enforced. One obtains
\begin{align}
g_+(\omega)&=
-\frac{\lambda_E^2}{6\omega_0^2}
 \biggl\{(\omega -2 \omega_0)  \text{Ei}\left(-\frac{\omega}{\omega_0}\right) +  
  (\omega +2\omega_0) e^{-\omega/\omega_0}\left(\ln \frac{\omega}{\omega_0}+\gamma_E\right)-2 \omega e^{-\omega/\omega_0}\biggr\}
\notag\\
&\quad + \frac{\e^{-\omega/\omega_0}}{2{ \omega_0}}\omega^2\biggl\{1 - \frac{1}{36\omega_0^2}(\lambda_E^2- \lambda_H^2)\biggr\},
\label{gplus-model1}
\end{align}
where $\text{Ei}(x)$ is the exponential integral.
For small momenta
\begin{align}
  g_+(\omega) &= \frac{\omega^2}{2{ \omega_0}}\biggl\{1 - \frac{\lambda_E^2}{6\omega_0^2} 
- \frac{(\lambda_E^2- \lambda_H^2)}{36\omega_0^2}\biggr\} + \mathcal{O}(\omega^3)\,
\notag\\&=
 \frac{9\omega^2}{16 \bar\Lambda}\biggl\{1 +\frac{(\lambda_E^2- \lambda_H^2)}{8\bar\Lambda^2}\biggr\} + \mathcal{O}(\omega^3)\,. 
\end{align}
The function $g_+(\omega)$ \eqref{gplus-model1} is plotted in Fig.~\ref{fig:g+model1} where we have taken $R =1/2$ \eqref{REH}.
It is interesting that despite a rather elaborate analytic expression in \eqref{gplus-model1} this function can be approximated 
with a very good accuracy by the simple expression
\begin{align}
   g_+(\omega) &\simeq \frac{3}{16\omega_0}\,\frac{3+4 R}{1+ 2 R}\, \omega^2 e^{-\omega/\omega_0}\,. 
\end{align}
Note that all EOM relations between the DAs are linear so that a more general model can be constructed 
as an arbitrary linear combination of the above expressions with different values of $\omega_0$.

For the leading-twist DA containing a large-momentum ``tail'' $\phi_+(\omega) \sim \frac{\ln \omega}{\omega}$ 
the definition of $g_+(\omega)$ in Eq.~\eqref{KKQT2} and the EOM relation \eqref{mod1-EOM} both have to be modified.
The problem is seen, e.g., using the expansions at $\mu s \ll 1$ 
\begin{align}
 \eta_+(s,\mu) = C_0  - \frac23 C_1 \bar\Lambda s  + \mathcal{O}(s^2)
\end{align}
with~\cite{Feldmann:2014ika}
\begin{align}
   C_0 &= 1+ \frac{\alpha_sC_F}{4\pi} \left(- 2 \ln^2(\mu s) + 2 \ln(\mu s) -2 - \frac{\pi^2}{12}\right) +\mathcal{O}(\alpha_s^2)\,,
 \nonumber\\
   C_1 &= 1+ \frac{\alpha_sC_F}{4\pi} \left(- 2 \ln^2(\mu s) + 2 \ln(\mu s) +\frac54 - \frac{\pi^2}{12}\right) +\mathcal{O}(\alpha_s^2)\,.
\label{small-s-eta+}
\end{align}
Using this expansion it is easy to check that the expression on the l.h.s. of Eq.~\eqref{mod1-EOM} 
acquires terms $\sim \alpha_s\,s\bar\Lambda\,\ln (\mu s)$
that do not match the assumed $\mathcal{O}(s^2)$ behavior of the twist-four contributions.  

Before a more satisfying solution is available, we suggest to use the low-momentum part of $\phi_+(\omega)$ only for 
the construction of the higher-twist DA models at a low reference scale.  
Alternatively, the sensitivity to large-momentum contributions can be removed by imposing EOM for the integrated coefficient functions 
(in $s$-space), cf.~\eqref{secondEOM1}.   
In this way one obtains the relation
\begin{align}
\frac{1}{\lambda_B} -  2\bar\Lambda \phi'_+(0)  =  -\frac{1}{9\omega_0^3} (2\lambda_E^2+\lambda_H^2)\,,
\end{align}
which can be used as a constraint on the twist-four parameters for a more general leading-twist DA model.

\begin{figure}[ht]
\begin{center}
\begin{picture}(300,270)(0,0)
\put(-5,0){\epsfxsize11cm\epsffile{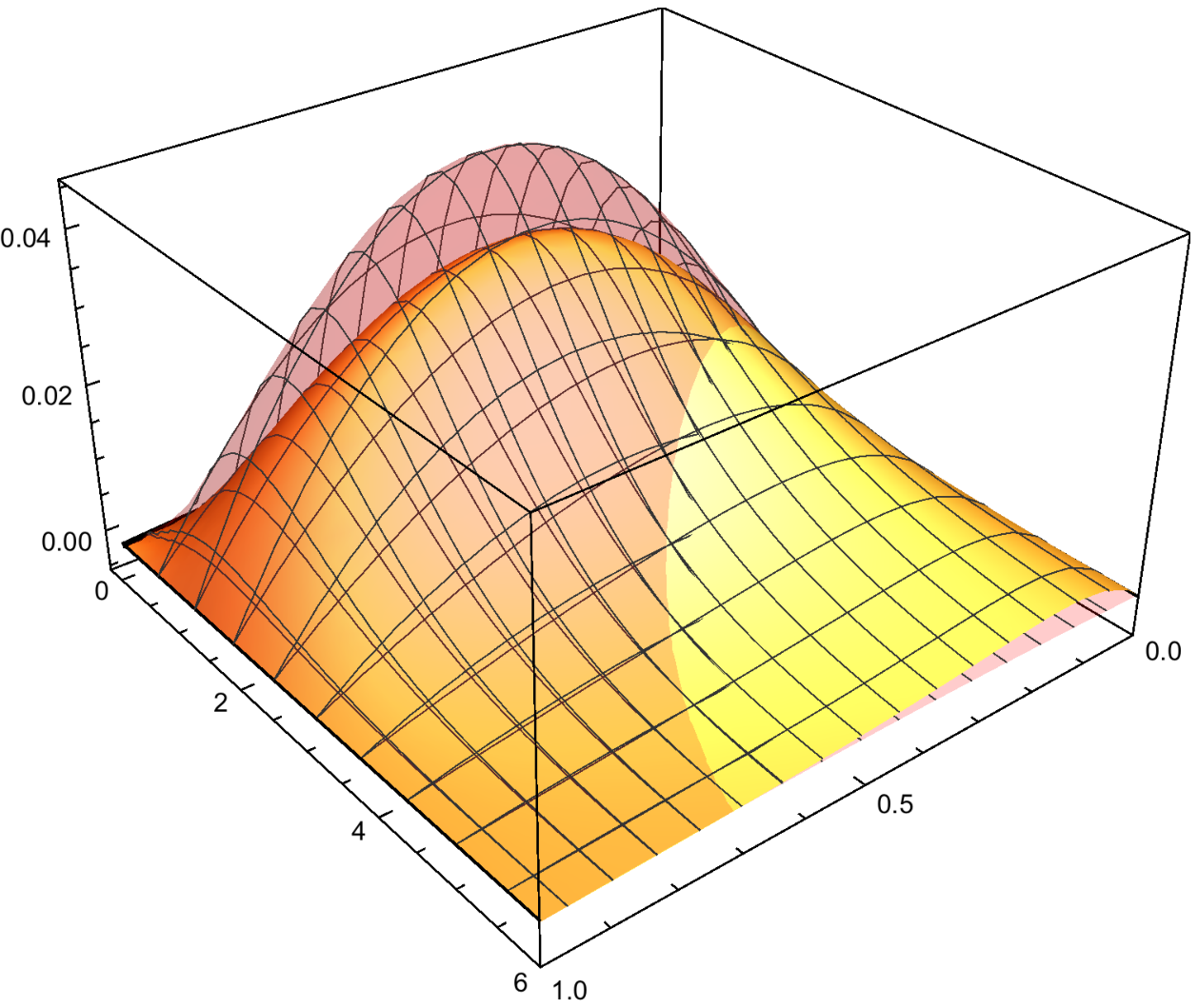}}
\put(45,55){\Large{$\dfrac{\omega}{2\omega_0}$}}
\put(235,40){\Large{$ u $}}
\put(200,205){{\Large $\psi_4+\tilde\psi_4$}}
\end{picture}
\\[-2mm]
\begin{picture}(300,270)(0,0)
\put(-5,0){\epsfxsize11cm\epsffile{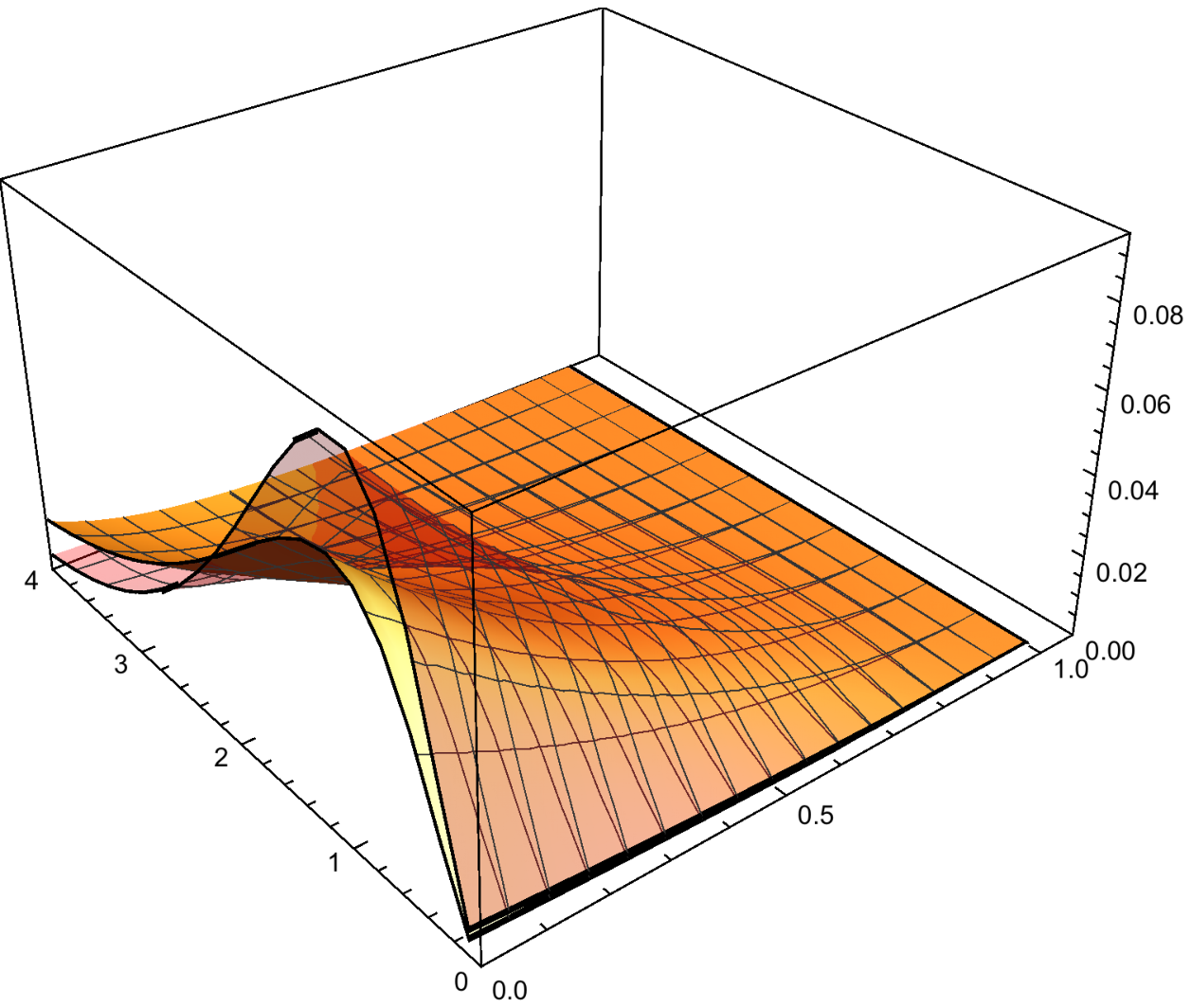}}
\put(85,8){\Large{$\dfrac{\omega}{2\omega_0}$}}
\put(265,38){\Large{$ u $}}
\put(220,185){{\Large $\phi_4$}}
\end{picture}
\end{center}
\caption{
Scale dependence of the DA ${\psi_4+\tilde\psi_4}$ (upper figure) and $\phi_4$ (lower figure) 
omitting the overall prefactor $(\lambda_E^2+\lambda_E^2)/\omega_0^2$.
The DAs at the scale $\mu=2.5$ GeV~(yellow) are overlaid with the input ones (rose) in~\eqref{model1-momspace} assuming $\mu_0=1$~GeV.}
\label{fig:scale-4-model1}
\end{figure}

The scale dependence of the twist-four DAs ${\psi_4+\tilde\psi_4}$ and $\phi_4$ omitting the overall
prefactor $(\lambda_E^2+\lambda_E^2)/\omega_0^2$ is shown in the upper and the lower figure in Fig.~\ref{fig:scale-4-model1},
respectively. The DAs at the scale $\mu=2.5$ GeV, shown in yellow, are overlaid with the input expressions~\eqref{model1-momspace}
where we assumed, for definiteness, $\mu_0=1$~GeV. In this plot we use the variables
\begin{align}
   \omega = \omega_1+\omega_2\,,\qquad \omega_1 = u\,\omega\,, \qquad \omega_2 = (1-u)\,\omega\,,
\end{align}
so that $\omega$ is the total momentum of the light degrees of freedom and $u$ is the fraction of the total momentum 
carried by the antiquark. Evolution has two effects. One of them is to suppress the higher-twist DAs at small momenta at larger scales
and create the large-momentum ``tails'' similar to the leading-twist DA. 
Another effect is that the three-particle DAs at higher scales are tilted towards 
a larger momentum fraction carried by the gluon so that the region $u\to 0$ is enhanced and the opposite region $u\to 1$ depleted.
This shift becomes more pronounced for larger values of the total momentum. This general pattern is seen in Fig.~\ref{fig:scale-4-model1}:  
the DAs at scale $\mu=2.5$~GeV are smaller than the input ones at $\mu=1$~GeV in the whole $u$ region for $\omega \lesssim 4\omega_0$
whereas for $\omega \gtrsim 4\omega_0 $ the DAs at $\mu=2.5$~GeV are larger than at $\mu=1$~GeV for an increasingly broad interval in $u$.   
Effects of the scale dependence on the twist-three DA $\phi_3$ are qualitatively similar and are discussed in detail in 
Ref.~\cite{Braun:2015pha}.

%

%
\subsection{Model II: Local duality}\label{subsec:Model-2}
%

Another class of models can be constructed using the local duality assumption, which is that the contribution of the B-meson
state to the correlation functions of the type 
\begin{align}
 &i \int d^4 y\, e^{-i\omega (vy)} \langle 0| T\{ \bar q(x) \Gamma_1 [x,0] h_v(0) \bar h_v(y)\Gamma_2 q(y)|0\rangle 
\notag\\
=&
  \langle 0| T\{ \bar q(x) \Gamma_1 [x,0] h_v(0) |\bar B(v)\rangle \frac{1}{\bar\Lambda-\omega} 
\langle \bar B(v)| \bar h_v(0)\Gamma_2 q(0)|0\rangle + \ldots   
\end{align}
is equal to the perturbative spectral density integrated over a certain energy interval (interval of duality).
Evaluating the simple quark loop and expanding in powers of $x^2$ one obtains in this way
\begin{align}
    \phi^{\rm LD}_+(\omega) &= \frac{3}{4\omega_0^3} \omega (2\omega_0-\omega) \theta (2\omega_0-\omega)\,,
\notag\\
    g_+^{\rm LD}(\omega) &=  \frac{3}{16\omega_0^3} \omega^2 (2\omega_0-\omega)^2 \theta (2\omega_0-\omega)\,,
\label{LD}
\end{align}  
with $\omega_0= (4/3) \bar\Lambda$.
This ``naive'' model for $g_+(\omega)$ can be used for a rough estimate of the size of the higher-twist effects.  

Selfconsistent models of this type can be constructed by including three-particle contributions. 
It turns out that the constraints due to the EOM~\eqref{KKQT4} require using the same 
threshold and power behavior $\sim (2\omega_0-\omega)^p$ for all DAs except for $\phi_3$ which has to be
one power lower. Two examples are given below.\\[2mm]

\noindent
{\bf Model IIA:~}Choosing $p=1$ we obtain 
\begin{align}
\phi_+(\omega,\mu_0)&=
\frac{3}{4\omega_0^3}\omega\,(2\omega_0-\omega)\,\theta(2\omega_0-\omega)\,,
\notag\\
\phi_-(\omega,\mu_0)&=
\frac{1}{8\omega_0^3}\left[3(2\omega_0-\omega)^2
-\frac{10(\lambda_E^2-\lambda_H^2)}{3\omega_0^2}\left(3\omega^2-6\omega\omega_0+2\omega_0^2\right)\right]\theta(2\omega_0-\omega),
\notag\\
\phi_3(\omega_1,\omega_2,\mu_0)&=
\frac{5(\lambda_E^2-\lambda_H^2)}{8\omega_0^5}\omega_1\omega_2^2\, \theta(2\omega_0-\bar\omega)\,,
\notag\\
\phi_4(\omega_1,\omega_2,\mu_0)&=
\frac{5(\lambda_E^2+\lambda_H^2)}{4\omega_0^5}\omega_2^2\,(\omega_0-\bar\omega/2) \theta(2\omega_0-\bar\omega)\,,
\notag\\
\psi_4(\omega_1,\omega_2,\mu_0)&=
\frac{5\lambda_E^2}{2\omega_0^5} \omega_1\omega_2\,(\omega_0-\bar\omega/2) \theta(2\omega_0-\bar\omega)\,,
\notag\\
\widetilde\psi_4(\omega_1,\omega_2,\mu_0)&=
\frac{5\lambda_H^2}{2\omega_0^5} \omega_1\omega_2\,(\omega_0-\bar\omega/2) \theta(2\omega_0-\bar\omega)\,,
\label{model2a}
\end{align} 
and
\begin{align}
g_+(\omega,\mu_0)&
=\frac{5\theta(2\omega_0-\omega)}{384\omega_0^5}\biggl\{\omega(2\omega_0-\omega)
\Big[8\lambda_E^2(\omega^2-4\omega\omega_0+2\omega_0^2)+\omega(2\omega_0-\omega)(2\lambda_H^2+9\omega_0^2)\Big]
\notag\\&\quad
+4\lambda_E^2\biggl[16\omega_0^3(\omega_0-\omega)\,\ln\Big(1-\frac{\omega}{2\omega_0}\Big)
+\omega^3(4\omega_0-\omega)\,\ln\Big(\frac{2\omega_0}{\omega}-1\Big)
\biggr]\biggr\},
\label{gplus-model2a}
\end{align}
where in the expressions for three-particle DAs  $\bar\omega=\omega_1+\omega_2$.
This set of DAs is consistent with EOM \eqref{KKQT4}, \eqref{secondEOM} provided the parameters satisfy the Grozin-Neubert 
constraints \eqref{GN}:
\begin{align}
 \omega_0=\frac32\lambda_B=\frac43\bar\Lambda\,, && 9\omega_0^2=40(2\lambda_E^2+\lambda_H^2)\,. 
\label{parfix:model2a}
\end{align}
The two-particle twist-4 DA $g_+(\omega)$ in Eq.~\eqref{gplus-model2a} can to a high accuracy be approximated by a simpler expression
\begin{align}
 g_+(\omega,\mu_0) \simeq  \frac{9}{128 \omega_0^3}\,\frac{12 R+7}{2R+1}\, \omega^2 (\omega_0-\omega/2)^2\theta(2\omega_0- \omega)\,.    
\end{align}
where $R$ is defined in \eqref{REH}.\\[2mm]

\noindent
{\bf Model IIB:~}Choosing $p=3$ we obtain instead 
\begin{align}
\phi_+(\omega,\mu_0)&
=\frac{5}{8\omega_0^5}\omega\,(2\omega_0-\omega)^3\,\theta(2\omega_0-\omega)\,,
\notag\\
\phi_-(\omega,\mu_0)&=
\frac{5(2\omega_0-\omega)^2}{192\omega_0^5}\left[6(2\omega_0-\omega)^2
-\frac{7(\lambda_E^2-\lambda_H^2)}{\omega_0^2}\left(15\omega^2-20\omega\omega_0+4\omega_0^2\right)\right] \theta(2\omega_0-\omega)\,,
\notag\\
\phi_3(\omega_1,\omega_2,\mu_0)&=
\frac{105(\lambda_E^2-\lambda_H^2)}{8\omega_0^7}\omega_1\omega_2^2\, (\omega_0-\bar\omega/2)^2\theta(2\omega_0-\bar\omega)\,,
\notag\\
\phi_4(\omega_1,\omega_2,\mu_0)&=
\frac{35(\lambda_E^2+\lambda_H^2)}{4\omega_0^7}\omega_2^2\,(\omega_0-\bar\omega/2)^3 \theta(2\omega_0-\bar\omega)\,,
\notag\\
\psi_4(\omega_1,\omega_2,\mu_0)&=
\frac{35\lambda_E^2}{2\omega_0^7} \omega_1\omega_2\,(\omega_0-\bar\omega/2)^3 \theta(2\omega_0-\bar\omega)\,,
\notag\\
\widetilde\psi_4(\omega_1,\omega_2,\mu_0)&=\frac{35\lambda_H^2}{2\omega_0^7} \omega_1\omega_2\,(\omega_0-\bar\omega/2)^3 \theta(2\omega_0-\bar\omega)\,,
\end{align}
and
\begin{align}
g_+(\omega,\mu_0)&=
\frac{\theta(2\omega_0-\omega)}{96\omega_0^7}\bigg\{\frac{5\omega^2}{16}\left(2\omega_0- \omega\right)^4(18\omega_0^2+7\lambda_H^2)
\notag\\&\quad 
-{7\lambda_E^2}\bigg[\frac{25}{48} \omega^6 -\frac{21}{4}\omega^5\omega_0
+ 22\omega^4 \omega_0^2-\frac{143}3\omega^3 \omega_0^3+45\omega^2 \omega_0^4 - 8 \omega \omega_0^5\bigg]
\notag\\&\quad
+\frac{7\lambda_E^2}{2}\bigg[
16 \omega_0^5 (2\omega_0-3\omega)\ln\Big(1-\frac{\omega}{2\omega_0}\Big)
\notag\\&\quad
+\omega^3 (4\omega_0- \omega)(\omega^2-5\omega\omega_0+10\omega_0^2)\ln\Big(\frac{2\omega_0}{\omega}-1\Big)
\bigg]\bigg\}.
\label{gplus-model2b}
\end{align}
where, as above, in the expressions for three-particle DAs  $\bar\omega=\omega_1+\omega_2$.
This set of DAs is consistent with EOM \eqref{KKQT4}, \eqref{secondEOM} provided the parameters satisfy the 
constraints \eqref{GN}:
\begin{align}
 \omega_0=\frac52\lambda_B=2\bar\Lambda\,,       && 3\omega_0^2=14(2\lambda_E^2+\lambda_H^2)\,. 
\label{parfix:model2b}
\end{align}
The two-particle twist-4 DA $g_+(\omega)$ in Eq.~\eqref{gplus-model2b} can to a high accuracy be approximated by a simpler expression
\begin{align}
 g_+(\omega,\mu_0) \simeq  \frac{5}{64\omega_0^5}\,\frac{20R+13}{2R+1}\, \omega^2 (\omega_0-\omega/2)^4\theta(2\omega_0- \omega)\,,    
\end{align}
where $R$ is defined in \eqref{REH}.

\begin{figure}[t]
\centerline{
\begin{picture}(210,140)(0,0)
\put(-5,0){\epsfxsize7.8cm\epsffile{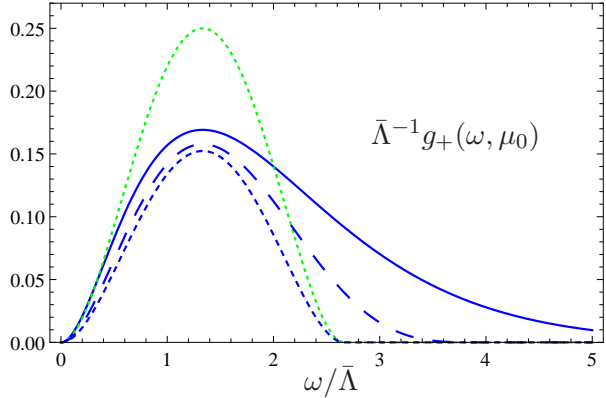}}
\put(105,-8){$\omega/\bar\Lambda$}
\put(130,85){$\bar\Lambda^{-1}g_+(\omega,\mu_0)$}
\end{picture}
}
\caption{
Four models for the two-particle twist-4 DA $g_+(\omega)$ described in the text: Model~I \eqref{gplus-model1} (solid),
Model~IIA \eqref{gplus-model2a} (short dashes), Model~IIB \eqref{gplus-model2b} (long dashes), and the ``naive'' local
duality model \eqref{LD} (green dots). 
For this plot we have taken $R = \lambda_E^2/\lambda_H^2 = 1/2$ \eqref{REH}.
}
\label{fig:g+}
\end{figure}

The four models for the two-particle twist-4 DA $g_+(\omega)$ described in the text are compared to each other in Fig.~\ref{fig:g+}.
Note that the model parameter $\omega_0$ has in each case to be adjusted to the same physical scale $\bar\Lambda$.
By construction the four considered models have different high-energy behavior but their difference at small momenta proves to be 
rather moderate. This is encouraging and allows one to hope that the model uncertainties in the 
power-suppressed contributions to $B$-meson decay form factors can be kept under control.    

%

%
\section{Summary}\label{sec:Summary}
%

Motivated by the challenge to control power-suppressed $1/m_B$ contributions to B-decays in final states with 
energetic light particles in the framework of QCD factorization, in this work we 
present a systematic study of three-particle higher-twist DAs of the B-meson which are main 
nonperturbative input in such analysis. 

We find eight independent three-particle DAs and classify them according to collinear twist and chirality which is 
related to their properties under conformal transformations. The twist-three three-particle DA $\phi_3$ and the related 
two-particle DA $\phi_-(\omega)$ has been studied in detail in Ref.~\cite{Braun:2015pha}. In this work we
concentrate on the three existing twist-four DAs. Our principal result, Eq.~\eqref{eq:t4final}, is the 
expansion of the DAs in terms of the eigenfunctions of the evolution equation and the calculation of the
corresponding anomalous dimension, Eq.~\eqref{gamma4}. We also derive a new relation between the two chiral-even DAs,
Eq.~\eqref{PhiPsirelation}, that is valid up to four-particle contributions of the type $\bar q GG h_v$ or $\bar q  q \bar q h_v$,
and a similar relation for the chiral-odd DAs that allows one to eliminate contributions with extra
transverse derivatives~\eqref{WWidentities}. 
We introduce two-particle higher-twist DAs that appear in the light-cone expansion of the heavy-light 
correlation functions following the approach of Kawamura \emph{et al.}~\cite{Kawamura:2001jm},
and discuss the corresponding EOM relations. We introduce several simple models for the higher-twist DAs  
with different large-energy behavior that satisfy all tree-level EOM constraints. We find that if the constraints due 
to EOM are enforced, model dependence of the low-momentum part of the two-particle twist-four DA $g_+(\omega)$
is rather mild, see Fig.~\ref{fig:g+}.  This is encouraging and allows one to hope that the twist-four uncertainties in the
predictions for physical observables in $B$-decays can be kept under control.

Several issues are not covered in this work and require further studies.
The first of them concerns the definition of two-particle twist-four (and higher) DAs that are auxiliary objects that 
arise via the light-cone expansion. The definition due to Kawamura \emph{et al.} that is employed in our work is sufficient
at the tree-level, but it has to be made more precise depending on the assumed power counting 
and particular factorization technique (e.g. SCET) beyond this accuracy. A related problem is that 
the EOM relation \eqref{KKQT4} between the DAs must be modified by $\mathcal{O}(\alpha_s)$ corrections as in the present 
form it is not consistent with the RG equations. It would be very important to find an alternative derivation of this 
relation that does not require considering off-light cone operators at the intermediate step.   

Another problem is of course that the theory of $1/m_B$ corrections to $B$-decays is at its infancy. It is well known that
the contributions of higher-twist DAs are plagued by end-point divergences so that the twist hierarchy is lost, in general.    
In recent years there had been some progress in this direction based on using dispersion 
relations~\cite{Khodjamirian:2006st,DeFazio:2007hw,Braun:2012kp,Wang:2015vgv,Wang:2016beq,Wang:2017jow}.
In this technique (light-cone sum rules) the hierarchy of higher-twist contributions is restored as an expansion in 
powers of the duality interval in the light hadron (photon) channel. Apart from (many) open theory issues, it remains to be seen, 
however, whether such techniques can provide one with phenomenologically relevant precision.

\section*{Acknowledgments}
Y. J. is grateful to Henry Lamm for useful comments.
The work of Y.J. and A.M. was supported by the DFG, grants BR 2021/7-1
and MO~1801/1-1, respectively.

\appendix
\section*{Appendices}
\addcontentsline{toc}{section}{Appendices}
\renewcommand{\theequation}{\Alph{section}.\arabic{equation}}
\renewcommand{\thetable}{\Alph{table}}
\setcounter{section}{0}
\setcounter{table}{0}

%
\section{Renormalization group equations}\label{app:RGequations}
%

The structure of the renormalization group equations for higher-twist operators is much simpler
in the spinor representation. To this end we consider the nonlocal (light-ray) twist-three
operator 
\begin{align}
 \mathbb{O}_3(z_1,z_2) & =\chi_+(z_1n)\bar f_{++}(z_2n)  h_+(0)\,,
\end{align}
and two doublets of twist-four operators corresponding to the quark and
gluon field of the same and opposite chirality
\begin{align}
  \mathbb{O}_4(z_1,z_2) & = \begin{pmatrix}\chi_-(z_1)f_{++}(z_2)h_v(0)\\ \chi_+(z_1)f_{+-}(z_2)h_v(0)\end{pmatrix},
&&
  \widebar{\mathbb{O}}_4(z_1,z_2)  =
\begin{pmatrix}\frac12\bar D_{-+}\chi_+(z_1)\bar f_{++}(z_2)h_v(0) \\ \chi_+(z_1)\bar f_{+-}(z_2)h_v(0)\end{pmatrix}.
\label{t4-doublet}
\end{align}
Matrix elements of these operators define the $B$-meson DAs of interest, see Sect.~\ref{subsec:spinor}. 
Note that  we introduce an extra operator with a transverse derivative acting on the
quark field. The corresponding matrix element defines a new DA, 
\begin{align}
  F_B(\mu)\, \Xi_4(z_1,z_2,\mu) &= \langle 0|\frac12 [\bar D_{\bar\mu\lambda}\chi_+](z_1n)\bar f_{++}(z_2n) \bar h_+(0) |\bar B(v)\rangle
\notag\\&=
 -\frac12  \langle 0| \bar q(z_1n) \stackrel{\leftarrow}{D}_\alpha^\perp [gG^{\alpha\beta}\gamma_5 + i g\widetilde G^{\alpha\beta}](z_2n) 
n_\beta \slashed{n} h_v(0)|\bar B(v)\rangle\,.
\label{def:Xi4}
\end{align}
We will show that this DA can be expressed in terms of the other ones using equations of motion (EOM) so
that the operator containing the transverse derivative can be eliminated. 
In this way, however, $SL(2)$ symmetry of the evolution equation becomes obscured.
It proves to be advantageous~\cite{Braun:2008ia,Braun:2009vc} to treat both operators as independent at 
the intermediate step and impose the EOM relations on the solutions.

The light-ray operators satisfy renormalization group equations
\begin{align}
\Big(\mu\frac{\partial}{\partial\mu}+\beta(\alpha_s)\frac{\partial}{\partial \alpha_s}
+\frac{\alpha_s}{2\pi}{\cal{H}}_3\Big)\mathbb{O}_3(z_1,z_2) &=0\,,
\notag\\
\Big(\mu\frac{\partial}{\partial\mu}+\beta(\alpha_s)\frac{\partial}{\partial \alpha_s}
+\frac{\alpha_s}{2\pi}{\cal{H}}_4\Big)\mathbb{O}_4(z_1,z_2) &=0\,,
\notag\\
\Big(\mu\frac{\partial}{\partial\mu}+\beta(\alpha_s)\frac{\partial}{\partial \alpha_s}
+\frac{\alpha_s}{2\pi}\widebar{\cal H}_4\Big) \widebar{\mathbb{O}}_4(z_1,z_2) &=0\,.
\end{align}
The evolution kernels $\mathcal{H}$ are integral operators acting on the light-cone coordinates of
the fields~\cite{Balitsky:1987bk} and, to one-loop accuracy,
can be written as sums of two-particle kernels which describe the interaction between the partons,
\begin{align}
\label{eq:kernels}
     {\cal H}&=  {\cal H}_{qh} + {\cal{H}}_{gh} +  {\cal{H}}_{qg} \,.
\end{align}
The twist-$4$ kernels, $\mathcal{H}_4$ and $\widebar{\mathcal{H}}_4$, are $2\times 2$ matrices, 
\begin{align}
\label{eq:hatHgeneral}
 {\cal{H}}_{qh}= \begin{pmatrix} H_{qh}^{11} & 0 \\ 0 &  H_{qh}^{22}\end{pmatrix},
&&
{\cal{H}}_{gh} = \begin{pmatrix} H_{gh}^{11} & 0 \\ 0 &  H_{gh}^{22}\end{pmatrix},
&&
{\cal{H}}_{qg}= \begin{pmatrix} H_{qg}^{11} &  H_{qg}^{12} \\  H_{qg}^{21}  &  H_{qg}^{22}\end{pmatrix}.
\end{align}
and similar for $\widebar{\cal H}$.
Explicit expressions for the two-particle kernels are collected in Appendix~\ref{app:kernels}.

It is well known that evolution equations for the light quark and gluon operators enjoy 
$SL(2,\mathbb{R})$ symmetry at one loop level. Each light field transforms according to a 
certain representation of the $SL(2,\mathbb{R})$ group which is defined by one number, the conformal spin $j$. 
The conformal spins of ``plus'' and ``minus'' quark fields $\chi_+, \chi_-$ are  $j=1$ and $j=1/2$. The
gluon fields $f(\bar f)_{++}$ and $f(\bar f)_{+-}$ transform according to the representation 
with $j=3/2$ and $j=1$, respectively, and the (holomorphic) transverse derivative of the quark field,
$\bar D_{-+}\chi_+$, has spin $j=3/2$. The $SL(2,\mathbb{R})$ (conformal) symmetry implies that the 
evolution kernels commute with the generators of symmetry transformations
\begin{align}
S^+_{j,z}=z^2\partial_z+2jz \,, \qquad S^0_{j,z}=z\partial_z+j\,, \qquad S^-_{j,z}=-\partial_z\,,
\end{align}
so that the two-particle light kernels can be written in terms of the corresponding quadratic Casimir
operators~\cite{Bukhvostov:1985rn}. This $SL(2)$-invariant representation (cf.~App.~\ref{app:kernels})
is very convenient for the further analysis. 

For the heavy-light operators the $SL(2,R)$ symmetry breaks down. However, the evolution equations 
remain to be invariant with respect to the special conformal transformations~\cite{Knodlseder:2011gc}. 
Hence, e.g., the kernel $\mathcal{H}_3$ commutes with
\begin{align}
\mathbb{Q}_1^{(3)} = iS^+_{(1,\frac32)}\,,
\end{align}
where $S^+_{(1,\frac32)}$ is a two-particle generator  
\begin{align}
   S^+_{(j_1,j_2)}= S^+_{(j_1,z_1)}+ S^+_{(j_2,z_2)}\,.
\end{align}
The corresponding generators for the twist-4 kernels are $2\times 2$ matrices,
\begin{align}
\label{eq:Q1}
\mathbb{Q}_1&=i \begin{pmatrix}S^{+}_{(\frac12,\frac32)}&0\\0&S^{+}_{(1,1)} \end{pmatrix},
\qquad
\mathbb{\widebar Q}_1=i
\begin{pmatrix} S^{+}_{(\frac32,\frac32)}&0\\0&S^{+}_{(1,1)}\end{pmatrix}, 
\qquad
[\mathbb{Q}_1,\mathcal{H}_4]  = [\widebar{\mathbb{Q}}_1,\widebar{\mathcal{H}}_4]=0\,.
\end{align}
Note that the conformal spins of up and down components in the  twist-$4$ doublets \eqref{t4-doublet} are different.

This residual symmetry is sufficient to solve the evolution equation for the two-particle 
leading-twist DA~\cite{Braun:2014owa}, but is not enough for three-particle DAs 
which are functions of two variables. 
Fortunately it turns out that the leading large-$N_c$ contributions to the higher-twist equations 
\begin{subequations}
\begin{align}
\label{eq:H3expansion}
{\cal H}_3=N_c\mathbb{H}_3+N_c^{-1}\delta\mathbb{H}_3\,,\\
\label{eq:hatHexpansion}
{\cal H}_4=N_c\mathbb{H}_4+N_c^{-1}\delta\mathbb{H}_4\,,\\
\label{eq:barHexpansion}
\widebar{\cal H}_4=N_c\mathbb{\widebar H}_4+N_c^{-1}\delta\mathbb{\widebar H}_4\,,
\end{align}
\end{subequations}
possess a hidden symmetry, so that we are able to construct another operator that commutes with
the evolution kernel. For the twist-three case, this operator was found in Ref.~\cite{Braun:2015pha}:   
\begin{align}
\mathbb{Q}_2^{(3)}&=\frac94 i S_{(\frac32,z_2)}^+ - iS_{(\frac32,z_2)}^+\big[S_{(\frac32,z_2)}^+ S_{(1,z_1)}^- 
+ S_{(\frac32,z_2)}^0 S_{(1,z_1)}^0\big]
\notag\\&\hspace*{2cm}
- iS_{(\frac32,z_2)}^0\big[S_{(1,z_1)}^0 S_{(\frac32,z_2)}^+-S_{(\frac32,z_2)}^0 S_{(1,z_1)}^+\big]\,,
\end{align}
such that
\begin{align}
  [\mathbb{Q}^{(3)}_1,\mathbb{Q}^{(3)}_2] = [\mathbb {Q}^{(3)}_1,\mathbb{H}_3] =  [\mathbb {Q}^{(3)}_2,\mathbb{H}_3] = 0\,.
\end{align}
The corresponding operators (``conserved charges'') for the twist-4 kernels take the form
(this is a new result):
\begin{align}
\mathbb{Q}_2=\frac i2\{J_{12}, J_{23}\}\,,
&&
\widebar{\mathbb{Q}}_2=\frac i2\{\bar J_{12}, \bar J_{23}\}\,,
\end{align}
where
\begin{align}
J_{12}&=\frac12\,\mathbb{I}+
\begin{pmatrix}
-\partial_2z_{12}\partial_1z_{12}+z_{12}\partial_1&~  1-\partial_2z_{12}
\\[2mm]
\partial_1z_{12}                               &~ -\partial_1\partial_2z_{12}^2
\end{pmatrix},
\qquad
J_{23}=
\begin{pmatrix}
-S^+_{(\frac32,z_2)}&0 \\ 0&-S^+_{(1,z_2)}
\end{pmatrix},
\end{align}
and
\begin{align}
\bar J_{12}&=\begin{pmatrix}
\frac1{z_{12}}\partial_1\partial_2z_{12}^3+\frac72&~~\frac{1}{z_{12}}(\partial_1\partial_2z_{12}^2+2)\\
-z_{12}&~~\partial_1\partial_2z_{12}^2+\frac12
\end{pmatrix},
\qquad
\bar J_{23}=\begin{pmatrix}
S^+_{(\frac32,z_2)}&0 \\ 0&S^+_{(1,z_2)}
\end{pmatrix}.
\end{align}
The operators $\mathbb{Q}_i, \widebar{\mathbb{Q}}_i$, $i=1,2$ satisfy the commutation relations:
 \begin{align}
 \label{eq:hatcomm}
 [\mathbb{Q}_i,\mathbb{Q}_j]=[\mathbb {Q}_i,\mathbb{H}_4]=0\,,
&&
 [\widebar{\mathbb Q}_i,\widebar{\mathbb Q}_j]=[\widebar{\mathbb Q}_i,\mathbb{\widebar H}_4]=0\, .
  \end{align}
The above expressions for $\mathbb{Q}_2$ and  $\widebar{\mathbb{Q}}_2$ can be derived and the commutation
relations verified most easily using the Quantum Inverse Scattering Method (QISM). 
This derivation will be presented elsewhere.

With this construction, the problem in question becomes mathematically equivalent to a quantum-mechanical 
system with two degrees of freedom, with $\mathbb{H}$ playing the role of the Hamiltonian and
$\mathbb{Q}_1$, $\mathbb{Q}_2$ the conserved charges. Thanks to commutativity all three operators  
share the same set of the eigenfunctions. Thus, instead of trying to find the eigenfunctions 
of $\mathbb{H}$ directly, one can construct eigenfunctions 
of the charges $\mathbb{Q}_1$, $\mathbb{Q}_2$ which are much simpler.

The complete set of twist-three eigenfunctions was obtained in~\cite{Braun:2015pha},
\begin{align}
 Y_{3}(s,x\,|\,\underline{z}) &=\frac{is^2}{z_1^2 z_2^3}\int_0^1 du\,u\bar  u\, e^{is(u/z_1+\bar u/z_2)}\,
{}_2F_1\left(\genfrac{}{}{0pt}{}{-\frac12-ix,-\frac12+ix}{2}\Big|-\frac{u}{\bar u}\right),
\notag\\
Y^{(0)}_3(s\,|\,\underline{z})&=  Y_{3}(s,x={i}/{2}\,|\,\underline{z}) =
\frac{is^2}{z_1^2 z_2^3}\int_0^1 du\, u\bar  u\,e^{is(u/z_1+\bar u/z_2)}\,.
\end{align}
They are labeled by two quantum numbers $s>0$ and $x\in \mathbb{R}$ 
related to the eigenvalues of the conserved charges,
\begin{align}
\mathbb{Q}_1^{(3)} Y_{3}(s,x|\,\underline{z})=s\,Y_{3}(s,x|\,\underline{z})\,, &&
\mathbb{Q}_2^{(3)} Y_{3}(s,x|\,\underline{z})=-s\, x^2\,Y_{3}(s,x|\,\underline{z})\,.
\end{align}
The charges $\mathbb{Q}_i^{(3)}$ are self-adjoint operators w.r.t the $SL(2)$-invariant scalar product
so that the eigenfunctions $Y_{3}(s,x\,|\,\underline{z})$ are mutually  orthogonal, see App.~\ref{app:scalarproduct}.

For the twist-four case, in addition to the two conserved charges
$\mathbb{Q}_1$ and $\mathbb{Q}_2$, one can construct an extra, "supplementary" charge $\mathbb{Q}_3$, such that
$[\mathbb{Q}_{{1,2}},\mathbb{Q}_3]=0$. The "supplementary" charges have a rather simple form
\begin{align}
\mathbb{Q}_3=i\begin{pmatrix} -S_{g,3/2}^+&z_1/z_2S^+_{g,1}
\\
z_2/z_1S^+_{q,1/2}& - S^+_{q,1}
\end{pmatrix}\,,
&&
\widebar{\mathbb{Q}}_3=i\begin{pmatrix}
S^+_{q,3/2}\quad &(S^0_{q,1}+1)(S^0_{g,1}+1)\\
z_1z_2\quad&S^+_{g,1}
\end{pmatrix}\,,
\end{align}
and are quite helpful for constructing of the eigenfunctions.

To this end we start with the following ansatz
\begin{align}
\Phi_k(\underline{z})= {z_1^{-2j_1} z_2^{-2j_2}} \int_0^1 du e^{is(u/z_1+\bar u/z_2)} \varphi_k(u)\,,
\end{align}
where $j_1$ and $j_2$ are the conformal spins of the light fields in the corresponding operator.
Such functions are, by construction, eigenfunctions of the first charge $\mathbb{Q}_1$ ($\widebar{\mathbb{Q}}_1$):
\begin{align}
\mathbb{Q}_1\begin{pmatrix}\Phi_1(\underline{z})\\\Phi_2(\underline{z})\end{pmatrix}
=s \begin{pmatrix}\Phi_1(\underline{z})\\\Phi_2(\underline{z})\end{pmatrix}.
\end{align} 
Requiring that the supplementary charge $\mathbb{Q}_3$ ($\widebar{\mathbb{Q}}_3$) is diagonalized yields
a simple relation between the ``upper'' and ``lower'' components,  $\varphi_1(u)$ and $\varphi_2(u)$,
after which the eigenvalue problem for $\mathbb{Q}_2$ ($\widebar{\mathbb{Q}}_2$) reduces to a second order
differential equation for, e.g., $\varphi_1(u)$. 
Selecting the solutions with the required analytic properties one obtains the following set of
eigenfunctions: \\
$\bullet$\hskip5mm Chiral operators (eigenfunctions of $\mathbb{Q}_i$-charges):
\begin{align}
\label{eq:Y-states-app}
\begin{pmatrix} Y_{4;1}^{(-)}\\ Y_{4;2}^{(-)}\end{pmatrix}(s,x\,|\,\underline{z}) &=
\frac{s^{3/2}}{z_1^2z_2^3}\int^1_0du\,\e^{is(u/z_1+\bar u/z_2)}\,
{}_2F_1\left(\genfrac{}{}{0pt}{}{\frac12-ix,\frac12+ix}{2}\Big|-\frac{u}{\bar u}\right)
\begin{pmatrix}u z_1\\[2mm] u z_2\end{pmatrix},
\notag\\[2mm]
\begin{pmatrix} Y_{4;1}^{(+)}\\ Y_{4;2}^{(+)}\end{pmatrix}(s,x\,|\,\underline{z}) &=
\frac{s^{3/2}}{z_1^2z_2^3}\int^1_0du\,\e^{is(u/z_1+\bar u/z_2)}\,
{}_2F_1\left(\genfrac{}{}{0pt}{}{ -ix, +ix}{1}\Big|-\frac{u}{\bar u}\right)
\begin{pmatrix}\bar u z_1 \\[1mm] -u z_2\end{pmatrix}.
\end{align}
$\bullet$\hskip 5mm Operators of different chirality  (eigenfunctions of $\widebar{\mathbb{Q}}_i$ charges):
\begin{align}
\label{app:Zfunctions}
\begin{pmatrix} Z_{4;1}^{(-)}\\ Z_{4;2}^{(-)}\end{pmatrix}(s,x\,|\,\underline{z}) &=
\frac{s^{3/2}}{z_1^3z_2^3}\int^1_0du\, {u^2}\,\e^{is(u/z_1+\bar u/z_2)}\,
{}_2F_1\left(\genfrac{}{}{0pt}{}{-ix, ix}{3}\Big|-\frac{u}{\bar u}\right)
\begin{pmatrix}is\,\bar u\\[2mm] z_1z_2\end{pmatrix},
\notag\\[2mm]
\begin{pmatrix} Z_{4;1}^{(+)}\\ Z_{4;2}^{(+)}\end{pmatrix}(s,x\,|\,\underline{z}) &=
\frac{s^{3/2}}{z_1^3z_2^3}\int^1_0du\,{u}{\bar u}\,\e^{is(u/z_1+\bar u/z_2)}\,
{}_2F_1\left(\genfrac{}{}{0pt}{}{-\frac12-ix,-\frac12+ix}{2}\Big|-\frac{u}{\bar u}\right)
\begin{pmatrix}-is\, u\\[2mm] z_1z_2\end{pmatrix}
\notag\\[2mm]
\begin{pmatrix} Z_{4;1}^{(0)}\\ Z_{4;2}^{(0)}\end{pmatrix}(s,x\,|\,\underline{z}) &=
\begin{pmatrix} Z_{4;1}^{(+)}\\ Z_{4;2}^{(+)}\end{pmatrix}(s,x=i/2\,|\,\underline{z}) ~=~
\frac{s^{3/2}}{z_1^3z_2^3}\int^1_0du \,u\bar u\,\e^{is(u/z_1+\bar u/z_2)}\,
\begin{pmatrix}-is\, u\\[2mm] z_1z_2\end{pmatrix}.
\end{align}
Both sets form a complete system with the respect of the $SL(2)$-invariant scalar product,
see App.~\ref{app:scalarproduct}.

Note that $Z_4^{(+)}$-functions are related to the twist-three eigenfunctions $Y_3$ as follows:
\begin{align}
Z_{4;1}^{+}(s,x|\underline{z}) = -\frac{i}{\sqrt{s}} \big(z_1\partial_{z_1}+2\big)\, Y_3(s,x|\underline{z})\,, &&
Z_{4;2}^{+}(s,x|\underline{z}) = -\frac{iz_2}{\sqrt{s}}\, Y_3(s,x|\underline{z})\,,
\label{ZYidentity}
\end{align}
and the same relation is valid between $Z_4^{(0)}$ and $Y_3^{(0)}$.

The last step is to calculate the eigenvalues of the Hamiltonians (anomalous dimensions):
\begin{align}
 N_c\mathbb{H}_4\begin{pmatrix} Y_{4;1}^{(\pm)}\\ Y_{4;2}^{(\pm)}\end{pmatrix} = \gamma_4^{\pm} \begin{pmatrix} Y_{4;1}^{(\pm)}\\ Y_{4;2}^{(\pm)}\end{pmatrix}, 
&&
 N_c \mathbb{\widebar H}_4\begin{pmatrix} Z_{4;1}^{(\pm)}\\ Z_{4;2}^{(\pm)}\end{pmatrix} = \widebar{\gamma}_4^{\pm} \begin{pmatrix} Z_{4;1}^{(\pm)}\\ Z_{4;2}^{(\pm)}\end{pmatrix}.
\end{align} 
This can most easily be done by comparing the large-$z_1,z_2$ asymptotic behavior on the both sides. In this way one obtains
\begin{align}
 \gamma_4^{(+)} &=  \widebar \gamma_4^{(-)} =  N_c[\ln (\mu s) +\gamma_E -5/4] + \gamma_4(x)\,,
\notag\\
 \gamma_4^{(-)} &=  \widebar \gamma_4^{(+)} =  N_c[\ln (\mu s) +\gamma_E -5/4] + \gamma_3(x)\,,
\end{align}
where $\gamma_3(x)$ and $\gamma_4(x)$ are defined in Eqs.~\eqref{gamma3} and \eqref{gamma4}, respectively.  

Expansion of the B-meson DAs over the eigenfunctions of the evolution equations reads
\begin{align}
\Phi_3(\underline{z},\mu) & = \int_0^\infty ds \Big[
\eta_3^{(0)}(s,\mu)\,{Y}_3^{(0)}(s\,| \underline{z}) + \frac12 \int_{-\infty}^\infty
dx\,\eta_3(s,x,\mu)\,{Y}_{3}(s,x\,|\,\underline{z})\Big], \label{Phi3-1}
\end{align}
and
\begin{align}
\begin{pmatrix}
-2\Phi_4(\underline{z},\mu)
\\[2mm]
[\Psi_4 + \widetilde\Psi_4](\underline{z},\mu)
\end{pmatrix}
& =- \int\limits^\infty_0{ds}\int\limits^\infty_{-\infty}dx\,
\biggl\{\eta_4^{(-)}(s,x,\mu)\,{Y}_{4}^{(-)}(s,x\,|\underline{z})
+ \eta_4^{(+)}(s,x,\mu)\,{Y}_{4}^{(+)}(s,x\,|\underline{z})\biggr\}\, ,
\notag\\[2mm]
\begin{pmatrix}
-2\Xi_4(\underline{z},\mu)
\\[2mm]
[\Psi_4 - \widetilde\Psi_4](\underline{z},\mu)
\end{pmatrix}
&=
 - \int\limits^\infty_0{ds}\int\limits^\infty_{-\infty}dx\,
\biggl\{\varkappa_4^{(-)}(s,x,\mu)\, {Z}_{4}^{(-)}(s,x\,|\underline{z})
+ \varkappa_4^{(+)}(s,x,\mu)\,{Z}_{4}^{(-)}(s,x\,|\underline{z})\biggr\}
\notag\\
&\quad + {2}\int\limits^\infty_0{ds}\,
\varkappa_4^{(0)}(s,\mu)\,{Z}_{4}^{(0)}(s\,|\underline{z})\,,
\label{eq:t4general}
\end{align}
where
\begin{align}
{Y}_{4}^{(\pm)} = \begin{pmatrix} {Y}_{4;1}^{(\pm)}\\ {Y}_{4;2}^{(\pm)}\end{pmatrix},
\qquad\qquad
{Z}_{4}^{(\pm)} = \begin{pmatrix} {Z}_{4;1}^{(\pm)}\\ {Z}_{4;2}^{(\pm)}\end{pmatrix}.
\end{align}
The coefficient functions $\eta_3$, $\eta_4^{(\pm)}$ and  $\varkappa_4^{(0,\pm)}$ have autonomous scale dependence
(up to $1/N_c^2$ corrections):
\begin{align}
 \eta_3(s,x,\mu) &= L^{\gamma_3(x)/\beta_0}R(s;\mu,\mu_0)\,\eta_3(s,x,\mu_0)\,,
\notag\\
   \eta_3^{(0)}(s,\mu) &= L^{N_c/\beta_0}R(s;\mu,\mu_0)\eta_3^{(0)}(s,\mu_0)\,,
\notag\\
   \eta_4^{(+)} (s,x,\mu) &= L^{\gamma_4(x)/\beta_0}R(s;\mu,\mu_0)\,  \eta_4^{(+)}(s,x,\mu_0)\,,
\notag\\
   \eta_4^{(-)} (s,x,\mu) &= L^{\gamma_3(x)/\beta_0}R(s;\mu,\mu_0)\,  \eta_4^{(-)}(s,x,\mu_0)\,,
\notag\\
   \varkappa_4^{(+)} (s,x,\mu) &= L^{\gamma_3(x)/\beta_0}R(s;\mu,\mu_0)\,  \varkappa_4^{(+)}(s,x,\mu_0)\,,
\notag\\
   \varkappa_4^{(-)} (s,x,\mu) &= L^{\gamma_4(x)/\beta_0}R(s;\mu,\mu_0)\,  \varkappa_4^{(-)}(s,x,\mu_0)\,,
\notag\\
   \varkappa_4^{(0)} (s,\mu) &= L^{N_c/\beta_0}R(s;\mu,\mu_0)  \varkappa_4^{(0)}(s,\mu_0)\,,
\label{scaledep:t34}
\end{align}
where $L = {\alpha_s(\mu)}/{\alpha_s(\mu_0)}$ and $R(s;\mu,\mu_0)$ is defined in Eq.~\eqref{Rfactor}.

The expressions in \eqref{eq:t4general} are valid for the most general case. We will show in App.~\ref{app:WW} that
neglecting contributions of four-particle (quasipartonic)~\cite{Bukhvostov:1985rn} operators of the 
type $\bar q GG h_v$ and $\bar q q\bar q h_v$,
the following relations hold:
\begin{align}
2\big(z_1\partial_{z_1}+1\big)\Phi_4(\underline{z}) &= \big(z_2\partial_{z_2}+2\big)\left[\Psi_4 (\underline{z})+\widetilde\Psi_4 (\underline{z})\right],
\notag\\
2z_1\,{\Xi_4}(\underline{z})&=
\big(z_2\partial_{z_2}+2\big)\left[\Psi_4 (\underline{z})-\widetilde\Psi_4 (\underline{z})\right]-2\Phi_3(\underline{z})\,.
\label{WWidentities}
\end{align}
Using the representation in Eq.~\eqref{eq:t4general} and taking into account Eq.~\eqref{ZYidentity} it is easy to show 
that the relations between the DAs in \eqref{WWidentities} imply the following relations between the coefficient 
functions (to the same accuracy): 
\begin{align}
 &\eta_4^{(-)}(s,x,\mu_) = 0\,,
\notag\\
  &\sqrt{s} \varkappa_4(s,x,\mu) = \eta_3(s,x,\mu)\,, 
\notag\\
  &\sqrt{s} \varkappa^{(0)}_4(s,\mu) = \eta^{(0)}_3(s,\mu)\,. 
\end{align}  
The resulting simplified expressions for the DAs are given in Eq.~\eqref{eq:t4final} in the main text.
We remind that quasipartonic four-particle operators do not mix with three-particle operators to the one-loop accuracy~\cite{Bukhvostov:1985rn}. 
For this reason neglecting such contributions is consistent with the scale dependence.

%
\section{Conformal scalar product}\label{app:scalarproduct}
%

Finding a suitable scalar product on the space of the B-meson  DAs 
is an auxiliary, although very useful, tool for solving the evolution equations.
The requirements for a scalar product are  the following:  i) the DAs under consideration have
to belong to the corresponding Hilbert space ii) the evolution kernels have to be self-adjoint operators.
It is not determined uniquely by the physics of the problem so that there is a certain freedom. 

The starting observation is that, first, support properties of the $B$-meson DAs in momentum space, $\omega>0$, imply
that they are analytic functions of the light parton coordinates $z$ in the lower half plane,
$\text{Im}\,z<0$ and, second,  the evolution kernels at one loop order are functions of the $SL(2,\mathbb{R})$ generators.
These  properties invite for using a formalism where $z$ is treated as a complex number and conformal symmetry of
the equations is implemented explicitly. Such a formalism is well known in mathematical literature.

One  defines the $SL(2,R)$ invariant scalar product for functions holomorphic in the lower complex
half-plane~\cite{Gelfand}
\begin{align}
\label{eq:SL2scalarproduct}
\vev{\Phi_1\big|\Phi_2}_j=\int_{\mathbb{C}_-}{\cal D}_jz\,\Phi^*_1(z)\,\Phi_2(z)\, ,
\end{align}
where the integration goes over  the lower half-plane $\mathbb{C}_-$ of the complex plane,
$\text{Im}\,z<0$, and the integration measure for spin $j$ is defined as 
$${\cal D}_jz=\frac{2j-1}\pi d^2z\,[i(z-\bar z)]^{2j-2}.$$ 
The scalar product~\eqref{eq:SL2scalarproduct} is invariant w.r.t. to the $SL(2,\mathbb{R})$
transformations~\cite{Gelfand}
\begin{align}
 \Phi(z) \mapsto \frac{1}{(cz+d)^{2j}} \Phi\left(\frac{az+b}{cz+d}\right)\,, \qquad ad-bc =1\,.
\end{align}
The generator of special conformal transformation $i S^{(j)}_+$ is self-adjoint w.r.t. this scalar product and its
eigenfunctions
\begin{align}
iS^{(j)}_+\, Q_s^{(j)}(z)=s\,Q_s^{(j)}(z)\,, && Q_s^{(j)}(z)=\frac{e^{-i\pi j}}{z^{2j}} e^{is/z}\,,
\end{align}
are orthogonal and form a complete set of functions in the Hilbert space~\footnote{This  is the so-called  Hilbert
space of holomorphic functions, see Ref.~\cite{Hall} for a review.}
 defined by Eq.~\eqref{eq:SL2scalarproduct}:
\begin{align}
&\langle{Q_s^{(j)}|Q_{s'}^{(j)}\rangle}_j=\frac{\Gamma(2j)}{s^{2j-1}}\,\delta(s-s')\,,
\\
&\frac1{\Gamma(2j)}\int_0^\infty ds \, s^{2j-1}\,Q_s^{(j)}(z)\,\overline{Q_s^{(j)}(z')} = \frac{e^{-i\pi j}}{(z-\bar z')^{2j}}\,.
\label{RK}
\end{align}
The expression on the r.h.s. of Eq.~\eqref{RK}
is the reproducing kernel (unit operator)~\cite{Hall},
i.e. for arbitrary function (holomorphic in the lower half-plane)
\begin{align}
\Psi(z)&=\frac{2j-1}\pi\int_{\mathbb{C}_-} \mathcal{D}_j z'\, \frac{e^{-i\pi j}}{(z-\bar z')^{2j}}\,\Psi(z')\,.
\end{align}
Exponential functions $e^{-i\omega z}$, $\omega >0$ form another complete orthogonal set
w.r.t. the same scalar product,
\begin{align}
\langle{e^{-i\omega z}|e^{-i\omega' z}\rangle}_j & ={\Gamma(2j)}\,{\omega^{1-2j}}\,\delta(\omega-\omega').
\label{ortho}
\end{align}
The momentum space DAs defined by the Fourier transform
\begin{align}
\Phi_j(z,\mu)=\int_0^\infty d\omega \, e^{-i\omega z }\,\phi_j(\omega,\mu)
\end{align}
can be found making use of \eqref{ortho} and the following relation:
\begin{align}
\langle e^{-i\omega z}|Q_s^{(j)}\rangle_j & = \Gamma(2j)\,(\omega s)^{1/2-j}\, J_{2j-1}(2\sqrt{\omega s})\,.
\end{align}
In this way one obtains
\begin{align}
\Phi_j(z,\mu)      &= \frac{1}{\Gamma(2j)} \int_0^\infty ds\, s^{2j-1} Q_s^{(j)}(z)\, \eta_j(s,\mu)\,,
\notag\\
\phi_j(\omega,\mu) &=\int_0^\infty ds\,\eta_j(s,\mu)\, (s \omega)^{j-1/2}\,J_{2j-1}(2\sqrt{\omega s})\,.
\end{align}
In particular for $j=1/2$ corresponding to the $B$-meson DA $\phi_-(\omega,\mu)$
the conformal expansion goes over Bessel functions $J_{0}(2\sqrt{\omega s})$ as
compared to $J_{1}(2\sqrt{\omega s})$ for the leading twist, in which case $j=1$.

The coefficient functions of DAs in the $s$-representation can be written as the scalar products as well,
e.g. for the leading twist
\begin{align}
 \eta_+(s,\mu) &= \langle  Q_s^{(1)} | \Phi_+\rangle\,.
\end{align}

The  $SL(2)$ invariant scalar product for the functions of two variables reads,
\begin{align}\label{scalar2}
 \vev{\Psi \big|\Phi}_{(j_1,j_2)}&=
\int_{\mathbb{C}_-}\!\!{\cal D}_{j_1}z_1\int_{\mathbb{C}_-}\!\!{\cal D}_{j_2}z_2\,\Psi^*(\underline{z})\,\Phi(\underline{z})
=
\Gamma(2j_1)\Gamma(2j_2)
\int^\infty_0\!\frac{d\omega_1}{\omega_1^{2j_1-1}}\frac{d\omega_2}{\omega_2^{2j_2-1}}\,\psi^*(\underline\omega)\phi(\underline\omega)
\end{align}
in position and momentum space representations, respectively.

The scalar product for the twist-$4$ doublets,
$\vec{\Phi}(\underline{z})=(\Phi_1(\underline{z}),\Phi_2(\underline{z}))$,  can be written in the form
\begin{align}
\vev{\overrightarrow{{\Phi}}|\overrightarrow{{\Psi}}} = \vev{\Phi_1|\Psi_1} + c \vev{\Phi_2|\Psi_2}\,,
\end{align}
where the two terms on the r.h.s. are defined by the scalar product~\eqref{scalar2} with the conformal spins matching
those of the corresponding --- up or down --- component of the doublet. The coefficient $c$ is fixed by the requirement that 
the Hamiltonian and conserved charges are self-adjoint operators. It is easiest to impose this condition on the 
``supplementary'' charges, $\mathbb{Q}_3 (\widebar{\mathbb{Q}}_3)$. For example, for the same-chirality doublet one has to 
require
\begin{align}
\langle \Phi_1|\mathbb{Q}_3^{12}\Psi_2\rangle_{(\frac12\frac32)}=c\,\langle \mathbb{Q}_3^{21}\Phi_1| \Psi_2\rangle_{(11)}\,,
\end{align}
where from it follows that $c=2$. In this way we obtain  (for all possible combinations of superscripts $\pm,0$ in the bra- and
ket-states)
\begin{align}
\vev{Y_4 (s,x)\big|Y'_4 (s',x')}=&\vev{Y_{4;1}(s,x)\big|Y'_{4;1}(s',x')}_{(\frac12,\frac32)}
+2\,\vev{Y_{4;2}(s,x)\big|Y'_{4;2}(s',x')}_{(1,1)}\,,
\notag\\
\vev{Z_4 (s,x)\big|Z'_4 (s',x')}=&\vev{Z_{4;1}(s,x)\big|Z'_{4;1}(s',x')}_{(\frac32,\frac32)}
+4\,\vev{Z_{4;2}(s,x)\big|Z'_{4;2}(s',x')}_{(1,1)}\,.
\end{align}

The normalization conditions of the twist-three and four eigenstates read:
\begin{align}
 \langle {Y}^{(0)}_{3}(s),{Y}^{(0)}_{3}(s') \rangle_{(1,\frac32)}&=\delta(s-s')\,, \notag\\
\langle {Y}_{3}(s,x),{Y}_{3}(s',x') \rangle_{(1,\frac32)}&=\delta(s-s') \delta(x-x') \frac{\coth\pi
x}{x(x^2+9/4)}\,,
 \notag\\
\vev{Y^{(-)}_4 (s,x)\big|Y^{(-)}_4 (s',x')}&=\delta(s-s')\delta(x-x')\frac{\coth\pi x}{x(x^2+1/4)}\, ,
\notag\\
\vev{Y^{(+)}_4 (s,x)\big|Y^{(+)}_4 (s',x')}&=\delta(s-s')\delta(x-x')\frac{\tanh\pi x}{x}\, ,
\notag\\
\vev{Z^{(-)}_4 (s,x)\big|Z^{(-)}_4 (s',x')}&=\delta(s-s')\delta(x-x')\frac{8\tanh\pi x}{x(x^2+1)(x^2+4)}\, ,
\notag\\
\vev{Z^{(+)}_4 (s,x)\big|Z^{(+)}_4 (s',x')}&=\delta(s-s')\delta(x-x')\frac{2\coth\pi x}{x(x^2+9/4)}\, ,
\notag\\
\vev{Z^{(0)}_4 (s)\big|Z^{(0)}_4 (s')}&=2\delta(s-s')\,.
\end{align}
Here it is assumed that $x>0$ and $x'>0$. This is sufficient because all eigenfunctions are symmetric
under reflection $x\to -x$. Scalar products for the pairs of the eigenfunctions with different superscripts vanish.

 The coefficient functions appearing in the expansion of the DAs in the eigenstates of the evolution
equation in \eqref{Phi3-1}, \eqref{eq:t4general} can be calculated as, for twist three,
\begin{align}
 \eta_3(s,x,\mu) &= \frac{x(x^2+9/4)}{\coth\pi x} \vev{Y_3(s,x)\big| \Phi_3}\,,
\qquad\qquad
 \eta^{(0)}_3(s,\mu) =  \vev{Y^{(0)}_3(s)\big| \Phi_3}\,,
\end{align}
and for twist four,
\begin{align}
\eta_4^{(-)}(s,x,\mu)&=
\frac{x(x^2+1/4)}{\coth\pi x}
\left[\vev{Y^{(-)}_{4;1}(s,x)\big|\Phi_4}_{(\frac12,\frac32)}-\vev{Y^{(-)}_{4;2}(s,x)\big|(\Psi_4+\widetilde\Psi_4)}_{(1,1)}\right],
\notag \\[2mm]
\eta_4^{(+)}(s,x,\mu)&= \frac{x}{\tanh\pi x}
\left[\vev{Y^{(+)}_{4;1}(s,x)\big|\Phi_4}_{(\frac12,\frac32)}
-\vev{Y^{(+)}_{4;2}(s,x)\big|(\Psi_4+\widetilde\Psi_4)}_{(1,1)}\right],
\notag\\[2mm]
\varkappa_4^{(-)}(s,x,\mu)&=\frac{x(x^2\!+\!1)(x^2\!+\!4)}{8\tanh\pi x}
\left[\vev{Z^{(-)}_{4;1}(s,x)\big|{\Xi}_4}_{(\frac32,\frac32)}-2\vev{Z^{(-)}_{4;2}(s,x)\big|(\Psi_4-\widetilde\Psi_4)}_{(1,1)}\right],
\notag\\[2mm]
\varkappa_4^{(+)}(s,x,\mu)&=\frac{x(x^2+9/4))}{2\coth\pi x}
\left[\vev{Z^{(+)}_{4;1}(s,x)\big|{\Xi}_4}_{(\frac32,\frac32)}-2\vev{Z^{(+)}_{4;2}(s,x)\big|(\Psi_4-\widetilde\Psi_4)}_{(1,1)}\right].
\end{align}
Note that $\eta_3^{(0)}(s,\mu)$ is related to the residue of $\eta_3(s,x,\mu)$ at imaginary $x=i/2$:
\begin{align}
 \eta_3(s,x,\mu)\Big|_{x\to i/2} = \frac{x (x^2+9/4)}{\coth \pi x} \eta_3^{(0)}(s,\mu)\Big|_{x\to i/2} =
 \frac{1}{x-i/2} \frac{i}{\pi} \eta_3^{(0)}(s,\mu) + \ldots
\end{align}

%
\section{Wandzura-Wilczek contributions in higher-twist operators}\label{app:WW}
%

In this Appendix we prove the following identities \eqref{WWidentities}:
\begin{align}
2\Big(z_1\partial_{z_1}+1\Big)\Phi_4(\underline{z}) &= \Big(z_2\partial_{z_2}+2\Big)\left[\Psi_4 (\underline{z})+\widetilde\Psi_4 (\underline{z})\right] +\ldots,
\notag\\
2z_1\,{\Xi_4}(\underline{z})&=
\Big(z_2\partial_{z_2}+2\Big)\left[\Psi_4 (\underline{z})-\widetilde\Psi_4 (\underline{z})\right]-2\Phi_3(\underline{z}) +\ldots\,.
\label{WWidentities-1}
\end{align}
where the ellipses stand for contributions of quasipartonic twist-four four-particle DAs of the type $\bar q GG h_v$ or $\bar q q\bar q h_v$.
We use a technique that has been developed, for light quarks, in 
Refs.~\cite{Braun:2008ia,Anikin:2013yoa} and is based on the two-component spinor formalism.
To this end it is convenient to relax the normalization condition Eq.~\eqref{norm-spinors} 
the auxiliary $\lambda$ and $\mu$ spinors, so that they can be treated as independent.

As a simpler example, let us first re-derive in this approach the familiar relation between the twist-three DAs, Eq.~\eqref{KKQT1}.
To start with, note that the matrix element involving plus components of the heavy quark and the 
antichiral light quark vanishes identically:
\begin{align} 
 \langle 0| \chi_+(z n)[zn,0] h_+(0) |\bar B\rangle =0.
\end{align}
Physics reason is that the quark helicities do not combine to zero. 
Formally, this matrix element vanishes because 
is not possible to construct a tensor with two ``undotted'' spinor indices 
$T_{\alpha\beta} = \langle 0| \chi_\alpha(z n)[zn,0] h_\beta(0) |\bar B\rangle$ such that 
$T_{++}=\lambda^\alpha\lambda^\beta T_{\alpha\beta}\neq 0$ from the two vectors 
$v_{\alpha\dot\alpha}$ and $n_{\alpha\dot\alpha}=\lambda_\alpha\bar\lambda_{\dot\alpha}$ 
at our disposal. Alternatively, this can be seen directly from the definition in Eq.~\eqref{def:two}.
 
The trick is to consider the derivative
\begin{align} 
 \mu^\alpha\frac{\partial}{\partial\lambda^\alpha}\langle 0| \chi_+(z n)[zn,0] h_+(0) |\bar B\rangle
\end{align} 
which, of course, must vanish as well.
Since $\chi_+=\lambda^\alpha \chi_\alpha$, $h_+=\lambda^\alpha h_\alpha$ and  $n_{\alpha\dot\alpha}=\lambda_\alpha\bar\lambda_{\dot\alpha}$,
applying the derivative $\mu\partial_\lambda$ 
we obtain three separate contributions that have to sum to zero:
\begin{align}
 0&=  \chi_-(z n)[zn,0] h_+(0) +\chi_+(z n)[zn,0] h_-(0) + 
      \frac12(\mu\sigma^\rho\bar\lambda)\frac{\partial}{\partial n^\rho}\chi_+(z n)[zn,0] h_+(0)\,,
\end{align}
where taking the matrix element $\langle 0|\ldots|\bar B\rangle$ is implied.

We stress that $n$ is a light-like vector. Nevertheless, taking the derivative $\partial/{\partial n^\rho}$ one can ignore 
the constraint $n^2=0$ and treat all four components of $n_\mu$ as independent ones. Indeed, let $F(n)$ be an arbitrary
function of $n_\mu$ defined  on the surface $n^2=0$ and extend it formally to 
$n^2\neq 0$,  $F(n)\mapsto F(n)|_{n^2=0} + n^2 F_1(n)$.%
\footnote{The difference of the present approach to the one used in \cite{Kawamura:2001jm} is that here we start 
with a matrix element of the renormalized light-ray operator which is finite by definition, whereas 
the light-cone expansion  $x^2\to 0$ in \cite{Kawamura:2001jm} produces singularities that have to be isolated 
in coefficient functions.}
Then
\begin{align}
\Big[(\mu\sigma^\rho\bar\lambda)\frac{\partial}{\partial n^\rho} n^2 F_1(n)\Big]_{n^2=0}=2(\mu n \bar\lambda)
F_1(n)=0\,,
\end{align}
so that the contribution of the added term vanishes: the answer does not depend on $F_1(n)$ which means that
the chosen extension beyond the light-cone surface does not matter.

Note that we have to differentiate both the (anti)quark field and the Wilson line:
\begin{align}
(\mu\sigma^\rho\bar\lambda)\frac{\partial}{\partial n^\rho}\chi_+(z n)[zn,0] h_+(0)
= 
 z \big[ \partial_{\mu\bar\lambda}\chi_+\big](z n)[zn,0] h_+(0)
+
\sigma^\rho_{\mu\bar\lambda} \chi_+(z n)\frac{\partial}{\partial n_\rho} [zn,0] h_+(0)\,,
\label{WWexample1}
\end{align}
where we use a shorthand notation
$\partial_{\mu\bar\lambda}= (\mu \partial \bar\lambda)=\mu^\alpha \partial_{\alpha\dot\alpha} \bar\lambda^{\dot\alpha}$. 
The derivative of the Wilson line gives
\begin{align}
 \frac{\partial}{\partial n^\rho} [zn,0] &=
iz A_\rho(zn)[zn,0] - i\int_0^z\! udu\, [zn,un] n^\sigma gG_{\sigma\rho}(un)[un,0]
\end{align}
and the term in $A_\rho(zn)$ combines with the derivative of the quark field in the first term in 
Eq.~\eqref{WWexample1} to produce a covariant derivative, $\partial_{\mu\bar\lambda} \mapsto D_{\mu\bar\lambda}$. 
This contribution can then be rewritten as a derivative over the light-cone coordinate 
using Fierz identity for Weil spinors $(u_1u_2)(v_1 v_2) = (u_1v_1)(u_2 v_2) - (u_1 v_2)(u_2 v_1)$
and EOM for the (massless) quark field $\bar D \chi=0$ (Dirac equation):
\begin{align}
\big[ D_{\mu\bar\lambda} \chi_+ \big](z n) & = [D_{\lambda\bar\lambda}] \chi_-(z n) +
(\lambda\mu) (\bar\lambda \bar  D \chi)(zn)=2 \partial_z  \chi_-(z n)\,. 
\end{align}
Rewriting the remaining term with the gluon strength tensor in spinor notation
\begin{align}
i \sigma^\rho_{\mu\bar\lambda}   \int_0^z u du\, gG_{\sigma\rho }(u n) n^{\sigma} 
=  ig (\lambda\mu) \int_0^z u du\, \bar f_{++}(un)\,,  
\end{align}
using that $h_{+} = - \bar h_-$, $ h_- =  \bar h_+$ \eqref{EOMheavy},
and collecting all contributions, we obtain the identity 
\begin{multline}
(z\partial_z+1)\langle 0| \chi_-(z n) \bar h_-(0)|\bar B(v)\rangle  =
\langle 0| \chi_+(z n) \bar h_+(0)|\bar B(v)\rangle
\\
-ig z^2 \int_0^1 u du \langle 0|\chi_+(z n) \bar f_{++}(uzn)  h_+(0)|\bar B(v)\rangle\,,
\end{multline}
where the gauge links are not shown for brevity,
which is equivalent to the relation in Eq.~\eqref{KKQT1}.

The same technique can be applied to three-particle operators. We start from the remark that 
the following matrix element vanishes:
\begin{align}
 \langle 0| \chi_+(z_1n)\bar f_{++}(z_2n) \bar h_+(0) |\bar B(v)\rangle = 0
\end{align}
and take a derivative  with respect to $\bar\lambda$:
\begin{eqnarray}
\lefteqn{
\hspace*{-1.5cm}\bar\mu^{\dot\alpha}\frac{\partial}{\partial\bar\lambda^{\dot\alpha}}\chi_+(z_1n)\bar f_{++}(z_2n) \bar h_+(0) =}
\nonumber\\ &=&
     \chi_+(z_1n)\bar f_{++}(z_2n) \bar h_-(0)
   + \frac12 z_1 [\bar D_{\bar\mu\lambda}\chi_+](z_1n)\bar f_{++}(z_2n) \bar h_+(0)
\notag\\ &&
   + 2 \chi_+(z_1n)\bar f_{+-}(z_2n) \bar h_+(0)
   + z_2 \frac12  \chi_+(z_1n) [\bar D_{\bar\mu\lambda} \bar f_{++}](z_2n) \bar h_+(0)+\ldots
\notag\\&=&
     - \chi_+(z_1n)\bar f_{++}(z_2n) h_+(0) 
     + \frac12 z_1 [\bar D_{\bar\mu\lambda}\chi_+](z_1n)\bar f_{++}(z_2n) \bar h_+(0) 
\notag\\&&
     + (z_2\partial_{z_2} + 2) \chi_+(z_1n)\bar f_{+-}(z_2n) \bar h_+(0) + \ldots\,,
\end{eqnarray}
where the ellipses stand for four-particle contributions (e.g. from derivatives of the Wilson lines) and EOM.
We also used that
\begin{align}
[\bar D_{\bar\mu\lambda} \bar f_{++}](z_2n)= D_{\lambda\bar\lambda} \bar f_{+-}(z_2n) + (\bar\mu\bar\lambda) (\lambda D \bar f)(z_2 n)
= 2 \partial_{z_2}\bar f_{+-}(z_2n) + (\bar\mu\bar\lambda) (\lambda D \bar f)(z_2 n)\,.
\end{align}
The last term can be rewritten as a bilinear  quark-antiquark operator using EOM for the gluon field 
$D_\beta^{~\dot\alpha} \bar f^a_{\dot\alpha\dot\beta} = g(\bar\psi_{\dot\beta} t^a \psi_\beta + \chi_\beta t^a \bar\chi_{\dot\beta})$
and gives rise to another four-particle contribution which we neglected.

Taking the matrix element $\langle 0|\ldots|\bar B\rangle$ and comparing with the definitions
of the DAs in Eqs.~\eqref{spinor23},~\eqref{spinor4},~\eqref{def:Xi4} we obtain
\begin{align}
2 z_1 \Xi_4(z_1,z_2,\mu) &= - 2 \Phi_3(z_1,z_2,\mu)  + (z_2\partial_{z_2} + 2)[\Psi_4-\widetilde\Psi_4] (z_1,z_2,\mu) +\ldots
\label{spinor-newEOM}
\end{align}
which is exactly the advertised second equation in \eqref{WWidentities-1}. 
This relation is exact up to contributions of four-particle quasipartonic twist-four operators.
It can be rewritten in a four-vector notation using 
\begin{align}
  z_1 \langle 0| \bar q(z_1)\!\derleft_{\alpha}\!gG^{\alpha\beta}(z_2)n_\beta\slashed{n}\gamma_5 h_v(0)  |\bar B(v)\rangle
  &=  F_B \Big[2 + z_1\partial_1 + z_2\partial_2\Big]  \Psi_4(\underline{z}) - F_B \Big[\Phi_3\! +\! \Phi_4\Big](\underline{z})\,,
\notag\\
  z_1 \langle 0| \bar q(z_1)\!\derleft_{\alpha}\!ig\widetilde{G}^{\alpha\beta}(z_2)n_\beta\slashed{n} h_v(0)  |\bar B(v)\rangle
  &=  F_B \Big[2 + z_1\partial_1 + z_2\partial_2\Big]  \widetilde{\Psi}_4(\underline{z}) + F_B \Big[{\Phi}_3\! -\! {\Phi}_4\Big](\underline{z})\,.
\label{newEOM1}
\end{align}

The first relation in Eq.~\eqref{WWidentities-1} can be derived in the same way starting from the matrix element
\begin{align}
\langle 0| \chi_+(z_1 n) f_{++}(z_2 n) h_-(0) |\bar B\rangle =0\,.
\end{align}
Taking the derivative over the auxiliary spinor $\mu\partial_\lambda$ and neglecting four-particle contributions one obtains the identity
\begin{align}
(z_1\partial_{z_1}+1)\langle 0| \chi_-(z_1 n) f_{++}(z_2 n) h_-(0) |\bar B\rangle +
(z_2\partial_{z_2}+2)\langle 0| \chi_+(z_1 n) f_{+-}(z_2 n) h_-(0) |\bar B\rangle =0\,,
\end{align}
which is equivalent to the first relation in~\eqref{WWidentities-1}.

%
\section{Two-particle evolution kernels}\label{app:kernels}
%
%
\subsection{Coordinate space representation}
%

\begin{align}
[ H^{11}_{qh}f](z_1)&=
-\frac1{N_c}\biggl\{\int^1_0\frac{d\alpha}\alpha\Big[f(z_1)-f(\bar\alpha z_1)\Big]
+\Big[\ln(i\mu z_1)-\frac54\Big]f(z_1)\biggr\},
\nonumber\\
[ H^{22}_{qh}f](z_1)&=
-\frac1{N_c}\biggl\{\int^1_0\frac{d\alpha}\alpha\Big[f(z_1)-\bar\alpha f(\bar\alpha z_1)\Big]
+\Big[\ln(i\mu z_1)-\frac54\Big]f(z_1)\biggr\},
\nonumber\\
[ H^{11}_{gh}f](z_2)&=\phantom{-}
N_c\biggl\{\int^1_0\frac{d\alpha}\alpha\Big[f(z_2)-\bar\alpha^2f(\bar\alpha z_2)\Big]
+\Big[\ln(i\mu z_2)-\frac12\Big]f(z_2)\biggr\},
\nonumber\\
[ H^{22}_{gh}f](z_2)&=\phantom{-}
N_c\biggl\{\int^1_0\frac{d\alpha}\alpha\Big[f(z_2)-\bar\alpha f(\bar\alpha z_2)\Big]
+\Big[\ln(i\mu z_2)-\frac12\Big]f(z_2)\biggr\},
\end{align}
\begin{align}
[ H^{11}_{qg}\varphi](z_1,z_2)&=
N_c\biggl\{\int^1_0\frac{d\alpha}\alpha\Big[2\varphi(z_1,z_2)-\varphi(z_{12}^\alpha,z_2)
-\bar\alpha^2\varphi(z_1,z_{21}^\alpha)\Big]-\frac34\varphi(z_1,z_2)\biggr\}
\notag\\&\quad
-\frac1{N_c}\int^1_0d\alpha\,\alpha\,\varphi(z_2,z_{12}^\alpha)\,,
\nonumber\\
[ H^{12}_{qg}\varphi](z_1,z_2)&=N_c\int^1_0d\alpha\,\varphi(z_{12}^\alpha,z_2)+\frac1{N_c}\int^1_0d\alpha\,\varphi(z_2,z^{\alpha}_{12})\,,
\nonumber\\
[ H^{21}_{qg}\varphi](z_1,z_2)&=N_c\int^1_0d\alpha\,\bar\alpha\,\varphi(z_1,z_{21}^\alpha)
+\frac1{N_c}\int^1_0d\alpha\,\bar\alpha\,\varphi(z_2,z^{\alpha}_{12})\,,
\nonumber\\
[ H^{22}_{qg}\varphi](z_1,z_2)&=N_c\biggl\{\int^1_0\frac{d\alpha}\alpha\Big[2\varphi(z_1,z_2)-\bar\alpha\varphi(z_{12}^\alpha,z_2)
 -\bar\alpha\varphi(z_1,z_{21}^\alpha)\Big]-\frac34\varphi(z_1,z_2)\biggr\},
\end{align}
and
\begin{align}
[\widebar H^{11}_{qh}f](z_1)&=
-\frac1{N_c}\biggl\{\int^1_0\frac{d\alpha}\alpha\Big[f(z_1)-\bar\alpha^2f(\bar\alpha z_1)\Big]+\Big[\ln(i\mu z_1)-\frac54\Big]f(z_1)\biggr\},
\nonumber\\
[\widebar H^{22}_{qh}f](z_1)&=
-\frac1{N_c}\biggl\{\int^1_0\frac{d\alpha}\alpha\Big[f(z_1)-\bar\alpha f(\bar\alpha z_1)\Big]
+\Big[\ln(i\mu z_1)-\frac54\Big]f(z_1)\biggr\},
\nonumber\\
[\widebar H^{11}_{gh}f](z_2)&=\phantom{-}
N_c\biggl\{\int^1_0\frac{d\alpha}\alpha\Big[f(z_2)-\bar\alpha^2f(\bar\alpha z_2)\Big]
+\Big[\ln(i\mu z_2)-\frac12\Big]f(z_2)\biggr\},
\nonumber\\
[\widebar H^{22}_{gh}f](z_2)&=\phantom{-}
N_c\biggl\{\int^1_0\frac{d\alpha}\alpha\Big[f(z_2)-\bar\alpha f(\bar\alpha z_2)\Big]+\Big[\ln(i\mu z_2)-\frac12\Big]f(z_2)\bigg\},
\nonumber\\
[\widebar H^{11}_{qg}\varphi](z_1,z_2)&=
N_c\biggl\{\int^1_0\frac{d\alpha}\alpha\big[2\varphi(z_1,z_2)
-\bar\alpha^2\,\varphi(z_{12}^\alpha,z_2)-\bar\alpha^2\,\varphi(z_1,z_{21}^\alpha)\big]
\nonumber\\&\quad
-2\int^1_0d\alpha\int^{\bar\alpha}_0d\beta\,\big(2\bar\alpha\bar\beta+\alpha\beta\big)\,
\varphi(z_{12}^\alpha,z_{21}^\beta)-\frac34\varphi(z_1,z_2)\bigg\}
\nonumber\\&\quad
-\frac6{N_c}\int^1_0d\alpha\int^1_{\bar\alpha}d\beta\,\bar\alpha\bar\beta\,\varphi(z_{12}^\alpha,z_{21}^\beta)\, ,\nonumber
\nonumber\\
[\widebar H^{12}_{qg}\varphi](z_1,z_2)&=
\frac{N_c}{z_{12}}\varphi_\pi(z_1,z_2)-\frac{2}{N_c z_{12}}\int^1_0\!d\alpha\int^1_{\bar\alpha}\!d\beta\,\Pi\,\varphi(z_{12}^\alpha,z_{21}^\beta)\, ,
\nonumber\\
[\widebar H^{21}_{qg}\varphi](z_1,z_2)&=
N_cz_{12}\biggl\{\int^1_0d\alpha\int^{\bar\alpha}_0d\beta\,(\bar\alpha\bar\beta+\alpha\beta)\varphi(z_{12}^\alpha,z_{21}^\beta)\bigg\}
\nonumber\\&\quad
+2\frac{z_{12}}{N_c}\int^1_0d\alpha\int^1_{\bar\alpha}d\beta\,\bar\alpha\bar\beta\,\varphi(z_{12}^\alpha,z_{21}^\beta)\, ,
\nonumber\\
[\widebar H^{22}_{qg}\varphi](z_1,z_2)&=
N_c\bigg\{\int^1_0\frac{d\alpha}\alpha\big[2\varphi(z_1,z_2)-\bar\alpha\,\varphi(z_{12}^\alpha,z_2)
-\bar\alpha\,\varphi(z_1,z_{21}^\alpha)\big]-\frac34\varphi(z_1,z_2)
\nonumber\\&\quad
-\int^1_0d\alpha\int^{\bar\alpha}_0d\beta\,\varphi(z_{12}^\alpha,z_{21}^\beta)
  +2\int^1_0d\alpha\,\alpha\bar\alpha\,\varphi(z_{12}^\alpha,z_{12}^\alpha)\bigg\}
\nonumber\\&\quad
-\frac2{N_c}\int^1_0d\alpha\,\alpha\bar\alpha\,\varphi(z^\alpha_{12},z^\alpha_{12})\, ,
\end{align}
where
\begin{align}
\Pi\,\varphi(z_1,z_2)&=\varphi(z_1,z_2)-6\int^1_0d\alpha\,\alpha\bar\alpha\,\varphi(z_{12}^\alpha,z_{12}^\alpha)
\end{align}
is the projector on the states with nonzero conformal spin.

\subsection{$SL(2)$-invariant representation} \label{app:SLkernels}
\label{}
It is also possible to present above integral forms of the evolution kernels in terms of $SL(2)$ transformation generators.
\begin{align}
[H^{11}_{qh}f](z_1)&=-\frac1{N_c}\big[\ln\left(i\mu S_{q,1/2}^+\right)-\psi(1)-5/4\big]\, ,\nonumber\\
[H^{22}_{qh}f](z_1)&=-\frac1{N_c}\big[\ln\left(i\mu S_{q,1}^+\right)-\psi(1)-5/4\big]\, ,\nonumber\\
[H^{11}_{gh}f](z_2)&=N_ c\big[\ln\left(i\mu S_{g,3/2}^+\right)-\psi(1)-1/2\big]\, , \nonumber\\
[H^{22}_{gh}f](z_2)&=N_c [\ln\left(i\mu S_{g,1}^+\right)-\psi(1)-1/2\big]\, ,\nonumber\\
[H^{11}_{qg}\varphi](z_1,z_2)&=N_c\Big[\psi\Big(J_{(\frac12,\frac32)}^{qg}+1\Big)+\psi\Big(J^{qg}_{(\frac12,\frac32)}-1\Big)-2\psi(1)-3/4\Big]\nonumber\\
&\quad+\frac1{N_c}(-1)^{J^{qg}_{(\frac12,\frac32)}+1}\frac{\Gamma\Big(J^{qg}_{(\frac12,\frac32)}-1\Big)}{\Gamma\Big(J^{qg}_{(\frac12,\frac32)}+1\Big)}\, ,\nonumber\\
[H^{12}_{qg}\varphi](z_1,z_2)&=N_c\frac{\Gamma\Big(J^{qg}_{(1,1)}-1\Big)}{\Gamma\Big(J_{(1,1)}^{qg}\Big)}
+\frac1{N_c}(-1)^{J_{(1,1)}^{qg}}\frac{\Gamma\left(J^{qg}_{(1,1)}-1\right)}{\Gamma\left(J^{qg}_{(1,1)}\right)}\, ,\nonumber\\
 [H^{21}_{qg}\varphi](z_1,z_2)&=N_c\frac{\Gamma\Big(J^{qg}_{(\frac12,\frac32)}\Big)}{\Gamma\Big(J_{(\frac12,\frac32)}^{qg}+1\Big)}
 +\frac1{N_c}(-1)^{J^{qg}_{(\frac12,\frac32)}}\frac{\Gamma\Big(J^{qg}_{(\frac12,\frac32)}\Big)}{\Gamma\Big(J^{qg}_{(\frac12,\frac32)}+1\Big)}\, ,\nonumber\\
  [H^{22}_{qg}\varphi](z_1,z_2)&=N_c\left[2\psi\left(J_{(1,1)}^{qg}\right)-2\psi(1)-3/4\right]\, ,
\end{align}
and
\begin{align}
[\widebar H^{11}_{qh}f](z_1)&=-\frac1{N_c}\big[\ln\left(i\mu S_{q,3/2}^+\right)-\psi(1)-5/4\big]\, ,\nonumber\\
[\widebar H^{22}_{qh}f](z_1)&=-\frac1{N_c}\big[\ln\left(i\mu S_{q,1}^+\right)-\psi(1)-5/4\big]\, ,\nonumber\\
[\widebar H^{11}_{gh}f](z_2)&=N_ c\big[\ln\left(i\mu S_{g,3/2}^+\right)-\psi(1)-1/2\big]\, , \nonumber\\
[\widebar H^{22}_{gh}f](z_2)&=N_c [\ln\left(i\mu S_{g,1}^+\right)-\psi(1)-1/2\big]\, ,
\nonumber\\
[\widebar H^{11}_{qg}\varphi](z_1,z_2)&=N_c\Big[\psi\Big(J^{qg}_{(\frac32,\frac32)}+2\Big)+\psi\Big(J^{qg}_{(\frac32,\frac32)}-2\Big)-2\psi(1)-\frac34\Big]\nonumber\\
&\quad+\frac{6}{N_c}(-1)^{J^{qg}_{(\frac32,\frac32)}}\frac{\Gamma\Big(J^{qg}_{(\frac32,\frac32)}-2\Big)}{\Gamma\Big(J^{qg}_{(\frac32,\frac32)}+2\Big)}\, ,\nonumber
\\
[\widebar H^{12}_{qg}\varphi](z_1,z_2)&=\frac1{z_{12}}\left[N_c-\frac{2(-1)^{J^{qg}_{(1,1)}}}{N_c}\frac{\Gamma\Big(J^{qg}_{(1,1)}-1\Big)}{\Gamma\Big(J^{qg}_{(1,1)}+1\Big)}\right]\Theta\left(J^{qg}_{(1,1)}>2\right)\, ,\nonumber\\
\end{align}
\begin{align}
[\widebar H^{21}_{qg}\varphi](z_1,z_2)&=\frac{N_cz_{12}}9\frac{\Gamma\Big(\frac13J^{gg}_{(\frac32,\frac32)}-2/3\Big)} {\Gamma\Big(\frac13J^{gg}_{(\frac32,\frac32)}+4/3\Big)}-\frac{2z_{12}(-1)^{J^{qg}_{(\frac32,\frac32)}}}{N_c}\frac{\Gamma\Big(J^{qg}_{(\frac32,\frac32)}-2\Big)} {\Gamma\Big(J^{qg}_{(\frac32,\frac32)}+2\Big)}\, ,\nonumber\\
[\widebar H^{22}_{qg}\varphi](z_1,z_2)&=N_c\Big[\psi\Big(J^{qg}_{(1,1)}+1\Big)+\psi\Big(J^{qg}_{(1,1)}-1\Big)-2\psi(1)-\frac34\Big]
-\frac{\delta_{J^{qg}_{(1,1)},2}}{3N_c}\, ,
\end{align}
where operator $J^{qg}$ is defined in terms of their corresponding quadratic Casimir operators 
$J^{qg}_{(j_1,j_2)}(J^{qg}_{(j_1,j_2)}-1)=(\vec S_{(q,j_1)}+\vec S_{(g,j_2)})^2$ and we have labeled the conformal spins 
of each $J^{qg}$ accordingly. $\Gamma(z)$ stands for the Euler-Gamma function while $\psi(z)$ is the $0$-th order polygamma function.

%

\end{document}